\documentclass[12pt,preprint]{aastex}

 \def\lsim{\lower 2pt \hbox{$\, \buildrel
{\scriptstyle <}\over {\scriptstyle \sim}\,$}}

\begin{document} \title{THE ROLE OF BEAM GEOMETRY IN POPULATION
STATISTICS AND PULSE PROFILES OF RADIO AND $\gamma$-RAY PULSARS}

\author{Peter L. Gonthier} \affil{Hope College, Department of Physics
and Engineering, 27 Graves Place \\ Holland, MI 49423-9000}
\email{gonthier@hope.edu}

\author{Robert Van Guilder} \affil{University Colorado at Denver,
Department of Physics, Denver, CO 80217} \email{rvanguilder@usa.net}

\and \author{Alice K. Harding} \affil{NASA Goddard Space Flight Center,
Laboratory for High Energy Astrophysics \\ Greenbelt, MD 20771}
\email{harding@twinkie.gsfc.nasa.gov}

\begin{abstract} We present results of a pulsar population synthesis
study that incorporates a number of recent developments and some
significant improvements over our previous study.  We have included the
results of the Parkes multi-beam pulsar survey in our select group of
nine radio surveys, doubling our sample of radio pulsars.  More
realistic geometries for the radio and $\gamma$-ray beams are included
in our Monte Carlo computer code that simulates the characteristics of
the Galactic population of radio and $\gamma$-ray pulsars.  We adopted
with some modifications the radio beam geometry of Arzoumanian, Chernoff
\& Cordes (2002).  For the $\gamma$-ray beam, we have assumed the slot
gap geometry described in the work of Muslimov \& Harding (2003). To
account for the shape of the distribution of radio pulsars in the $\dot
P-P$ diagram, we continue to find that decay of the magnetic field on a
timescale of 2.8 Myr is needed. With all nine surveys, our model
predicts that EGRET should have seen 7 radio-quiet (below the
sensitivity of these radio surveys) and 19 radio-loud $\gamma$-ray
pulsars. AGILE (nominal sensitivity map) is expected to detect 13
radio-quiet and 37 radio-loud $\gamma$-ray pulsars, while GLAST,
with greater sensitivity is expected to detect 276 radio-quiet and
344 radio-loud $\gamma$-ray pulsars. When the Parkes multi-beam
pulsar survey is excluded, the ratio of radio-loud to radio-quiet
$\gamma$-ray pulsars decreases, especially for GLAST.  The decrease
for EGRET is 45\%, implying that some fraction of EGRET
unidentified sources are radio-loud $\gamma$-ray pulsars. In the radio
geometry adopted, short period pulsars are core dominated.  Unlike the EGRET
$\gamma$-ray pulsars, our model predicts that when two $\gamma$-ray
peaks appear in the pulse profile, a dominant radio core peak appears in
between the $\gamma$-ray peaks.  Our findings suggest that further
improvements are required in describing both the radio and $\gamma$-ray
geometries. \end{abstract}




\section{Introduction} Rotation-powered pulsars are the brightest class
of $\gamma$-ray sources detected by the Compton $\gamma$-Ray Observatory
(CGRO).   The high-energy telescope EGRET made firm detections of pulsed
$\gamma$-ray emission from five known radio pulsars (Thompson 2001), and
possible detections from several others (Kanbach 2002).  In addition,
the high-energy pulsar Geminga may be radio-quiet, or at least radio-weak.  EGRET also discovered more than 200 $\gamma$-ray sources (Hartman
et al. 1999), most of which are still unidentified.  However, several
dozen new radio pulsars, out of more than 600 discovered since the end
of the CGRO mission by the recent Parkes multi-beam pulsar survey
(PMBPS) (Manchester et al. 2001) or in deep targeted observations
(Lorimer 2003), lie within the error circles of EGRET sources.  Although
many of these are young, energetic pulsars, their identification as
$\gamma$-ray pulsars must await observation with the next $\gamma$-ray
telescopes, AGILE and GLAST.

In the meantime, population synthesis studies of radio and $\gamma$-ray
pulsars can predict the number of radio-loud and radio-quiet
$\gamma$-ray pulsar detections expected by different telescopes,
assuming different models for radio and $\gamma$-ray emission.  Even
though current radio and $\gamma$-ray emission models have a number of
outstanding uncertainties, the results of such studies can provide quite
sensitive discrimination between models.  In particular, polar cap and
outer gap models of $\gamma$-ray emission make very different
predictions of the number of radio-loud and radio-quiet $\gamma$-ray
pulsars.  Polar cap models (e.g. Daugherty \& Harding 1996), where the
high-energy emission region is located on the same open field lines as
the radio emission, expect a large overlap in the radio and $\gamma$-ray
emission beams and thus  a higher ratio of radio-loud to radio-quiet
$\gamma$-ray pulsars.  On the other hand, outer gap models predict a
smaller overlap between $\gamma$-ray and radio emission beams, because
the high-energy and visible radio emission originate from opposite poles
(Romani \& Yadigaroglu 1995, Cheng et al. 2000), thus predicting more
radio-quiet than radio-loud $\gamma$-ray pulsars. Thus, population
synthesis can also address the question of how many EGRET unidentified
sources are radio pulsars.

Results of our initial study of pulsars in the Galaxy were presented by
Gonthier et al. (2002).  In this work, we evolved neutron stars from
birth distributions in space, magnetic field strength, period and kick
velocity, in the Galactic potential to simulate the population of radio
pulsars detected in eight surveys of the Princeton catalog (Taylor,
Manchester \& Lyne 1993).  A very simple model of radio and $\gamma$-ray
beams was assumed, in which both were aligned with solid angle of 1 sr.
Radio luminosity was assigned using the model of Narayan \& Ostriker
(1990) and $\gamma$-ray luminosity from the polar cap model of Zhang \&
Harding (2000).  We found that agreement of the distribution of
simulated radio pulsars with the observed distribution was significantly
improved by assuming decay of the neutron star surface magnetic field on
a time scale of 5 Myr.  With these assumptions, EGRET should have
detected 9 radio-loud and 2 radio-quiet $\gamma$-ray pulsars, and GLAST
should detect 90 radio-loud and 101 radio-quiet pulsars (9 detected as
pulsed sources).  Because the radio and $\gamma$-ray beam apertures were
assumed to be identical, Òradio-quietÓ $\gamma$-ray pulsars were those
whose radio emission is too weak to be detected by the selected radio
surveys.

There have been a number of new developments in both radio pulsar
observation and analysis, and in $\gamma$-ray pulsar theory since we
completed our initial population study.  The PMBPS (Manchester et al.
2001, Morris et al. 2002, Kramer et al. 2003) is nearly complete and has
more than doubled the number of radio pulsars with measured period
derivatives from 445 to nearly 1300.  Determination of pulsar distances
from dispersion measure has been greatly improved with the development
of a new model by Cordes \& Lazio (2002).  New radio luminosity and beam
models have been developed by Arzoumanian, Chernoff \& Cordes (2002),
which describe core and conal components of the emission and their
dependence on period and frequency.  Arzoumanian, Chernoff \& Cordes
(2002) have also derived a new two-component distribution of radio
pulsar space velocities.  A new polar cap $\gamma$-ray emission model
has been developed by Muslimov \& Harding (2003), in which radiation
from pair cascades at high altitude along the edge of a Ôslot gapÕ forms
a wide hollow cone of emission.  In addition, the solid angle as
well as the luminosity of the $\gamma$-ray beam is described in this
model.

This paper presents results of an expanded and updated pulsar population
synthesis study that includes all of the above recent developments, as
well as improved $\gamma$-ray sensitivity maps.  By incorporating
independent models for the radio and $\gamma$-ray beam geometry, we are
now able to investigate how the beam geometry affects the observable
characteristics of radio-loud and radio-quiet $\gamma$-ray pulsar
populations.  We are also able to address the question of how many EGRET
unidentified sources are expected to be radio-loud or radio-quiet
$\gamma$-ray pulsars in the polar cap model.  Of particular interest is
the issue of how many of the new Parkes radio pulsars in EGRET error
circles are counterpart $\gamma$-ray pulsars.  In addition, we can make
more accurate estimates of the numbers of radio-loud or radio-quiet
$\gamma$-ray pulsars detectable by the AGILE and GLAST telescopes.

\section{Radio Emission Geometry}

We have adopted the geometry model for the radio emission beams as
presented by Arzoumanian, Chernoff \& Cordes (2002) (ACC from now on)
with some slight modifications.  We have assumed a core and a single
conal component described by Gaussians with characteristic widths as
follows

\begin{eqnarray} \rho _{core}&=&1.5^\circ P^{-0.5},\ {\rm and} \nonumber
\\ \rho _{cone}&=&5.2^\circ \left( {1+{{66} \over \nu }} \right)
P^{-0.5} \label{eq:rhos} \end{eqnarray}

\noindent where the period, $P$, is in seconds and the frequency, $\nu$,
is in MHz.  The characteristic core width is the width of the core beam
at $1/e$ of the peak intensity.  We have incorporated the
radius-to-frequency mapping in the conal width developed by Mitra \&
Deshpande (1999); although they introduce elliptical shapes to the conal
geometry, they find no compelling reason to abandon circular beams.  The
coefficient of $5.2^\circ$ above is chosen to give the same width at 400
MHz as in ACC.

For each simulated pulsar, the pulse profile is binned into 500 bins of
the phase angle, $\phi$, ranging from $-\pi$ to $\pi$.  Each bin is
assigned a flux, $s(\phi,\nu)$, consisting of the sum of the flux
contributions from the core and cone components given by

\begin{eqnarray}
s(\phi,\nu)&=& S_{core}(\theta,\nu) + S_{cone}(\theta,\nu) \\ \nonumber
S_i(\theta,\nu)
&=&-{{\alpha_i+1}\over \nu} \left( {\nu\over {50\ {\rm MHz}
}}\right)^{\alpha_i+1} f_i(\theta) {L_i\over d^2}, 
\label{eq:flux1}
\end{eqnarray}

\noindent where $i$ indicates core or cone, $\alpha_i$ is the spectral
index ($S_i \propto\nu^{\alpha_i}$ with $\alpha_i < 0$), $\theta$ is the
polar angle to the magnetic axis, $f_i(\theta)$ is the angular
distribution of the component flux, $L_i$ is the component luminosity
and $d$ is the distance to the pulsar.  The relationship between the
phase angle, $\phi$, and the polar angle, $\theta$, depends on the
viewing geometry given by the expression

\begin{equation} \cos \theta =\sin \alpha \sin \zeta \cos \phi +\cos
\alpha \cos \zeta . \label{eq:theta} \end{equation}

\noindent During the simulation, the magnetic inclination angle,
$\alpha$, and the observer's line of sight angle, $\zeta$, are chosen
randomly between zero and $\pi/2$, accounting for emission from
both poles.  The difference between these two angles defines the impact
angle, $\beta = \zeta - \alpha$.

We assume that the spectrum of each component has a low frequency cutoff
of 50 MHz and can be modeled by a single power law, 
with spectral indices of $\alpha_{core}=-2.1$ and
$\alpha_{cone}=-1.6$. It is general agreed that the spectra of cores
are steeper that those of cones especially for short-period pulsars
(Lyne \& Manchester 1988 and Rankin 1983).  As discussed later in
section 5, we have assumed constant spectral indices with a difference
of 0.5 between the core and cone indices.

The angular distributions of the core and conal components are given by
the Gaussians

\begin{eqnarray} f_{core}(\theta) &=&{1\over
{\Omega_{core}}}e^{-\theta^2/\rho^2_{core}},\ {\rm and} \\ \nonumber
f_{cone}(\theta)
&=&{1\over {\Omega_{cone}}} e^{-(\theta - \bar\theta )^2/w^2_e}.
\label{eq:flux2} \end{eqnarray}

\noindent The solid angles for each of the components are chosen to
normalize the Gaussian distributions describing the angular distribution
of the flux in equation (\ref{eq:flux2}) when integrating over the polar
angle, $\theta$, the azimuthal angle, $\phi$, and are given by the
approximate expressions

\begin{eqnarray} \Omega_{core}&=&\pi \rho _{core}^2,\ {\rm and}\\
\nonumber \Omega_{cone}&=&2\pi ^{3/2}w_e\bar \theta \sim 0.8\pi \rho
_{cone}^2. \label{eq:omegas} \end{eqnarray} \noindent The width and
radius of the annulus of the conal beam are given by

\begin{eqnarray} 
w_e&=&{{\rho _{cone}} \over {4\sqrt {\ln 2}}},\ {\rm
and} \\ \nonumber 
\bar\theta &=&0.75\rho_{cone}=3.9^\circ \left(
1+{66\over \nu}\right)P^{-0.5}. 
\label{eq:widths} 
\end{eqnarray}

\noindent These expression differ slightly
from the ones used by Arzoumanian (private communication) given by the
following forms:

\begin{eqnarray} 
\Omega_{core}^{ACC}&=&{{\pi \rho _{core}^2} \over {\ln
2}}\sim 1.4\pi \rho _{core}^2,\ {\rm and}\\ \nonumber
\Omega_{cone}^{ACC}&=&{{2\pi \rho _{cone}^2} \over {5\ln 2}}\sim 0.58\pi \rho
_{cone}^2. 
\label{eq:omgasz} 
\end{eqnarray}

\noindent With the above definition of $w_e$ when $\theta=\rho_{cone}$,
$\rho_{cone}$ represents the radius of the cone at half power.  The
half-angle coefficient of $3.9^\circ$ in equation (6)
is the angle where the conal intensity peaks.  This coefficient
corresponds to that of the middle cone of the three conal components discussed
by Mitra \& Deshpande (1999). The phase angle, $\phi$, is varied between
$-\pi$ and $\pi$ and divided into 500 bins with the flux contributions
of core and conal components evaluated and summed.  In order to see if
the simulated pulsar is ÒdetectedÓ, the averaged flux, $S_{ave}$, of the
pulse profile is then compared to the flux threshold, $S_{min}$, of each
survey at its corresponding frequency. If detected, the pulsar is
flagged as radio-loud otherwise it is radio-quiet.

In the ACC study, pulsar surveys were all selected near a frequency of
400 MHz; hence they had no need to introduce any frequency dependence to
the fluxes of the core and conal components.  An important assumption
made in ACC is that the ratio of the core to cone flux is given by
$r={20\over 3}P^{-1}$.  In the present study, we have two groups of
pulsar surveys.  One group has frequencies near 400 MHz, while the other
group has frequencies near 1400 MHz. Therefore, we have had to introduce the
above frequency dependence to the spectra of the core and conal
components with a ratio of the core to cone peak flux given by

\begin{equation} r = 10 P^{-1}\left( {{\alpha_{core}+1}\over
{\alpha_{cone}+1}}\right) \left( {\nu\over {400\ {\rm MHz}}}
\right)^{\alpha_{core}-\alpha_{cone}}, \label{eq:coretocone}
\end{equation}

\noindent where this ratio is similar to the ratio used in ACC at a
constant frequency of 400 MHz.  With this ratio, short period pulsars
will have their radio fluxes dominated by the core component with a weak
conal component depending on the viewing geometry.

The luminosities of the core and cone components are expressed by

\begin{eqnarray} L_{cone}&=&{L \over {\left( {1+r/r_o} \right)}},\ {\rm
and}\cr L_{core}&=&{L \over {\left( {1+r_o/r} \right)}},\ {\rm where}\cr
r_o&=&{{\Omega _{cone}} \over {\Omega _{core}}}\left( {{\alpha_{core}+1}
\over {\alpha_{cone}+1}}\right) \left({\nu\over {50\ {\rm
MHz}}}\right)^{\alpha_{core}-\alpha_{cone}}, \label{eq:lumos}
\end{eqnarray}

\noindent where $r/r_o$ is the ratio of core to cone luminosities, and
$L$ is the total luminosity given by

\begin{equation} L=3.4\times 10^{10}P^{-1.3}\dot P^{0.4}\ ({\rm mJy\cdot
kpc^2\cdot MHz}). \label{eq:lumo} \end{equation}

\noindent This luminosity is reduced by a factor of 60 from the one used
in ACC, as discussed in section 5. Under this assumption, radio pulsars
are believed to be standard candles with well-defined luminosities in
terms of only the period and period derivative whose exponents in the
expression above for $L$ come from parameters in Table 1 of ACC for the
first model assuming a braking index $n=3$ with two velocity components.
There is no dithering of the luminosity as in the case of Narayan \&
Ostriker (1990).  Here the random viewing geometry accounts completely
for the required dithering when the beam and viewing geometries are not
included.  However as discussed later in the text in section 5, in order
to obtain a resonable birth rate and adequate agreement between the
distributions of the distance, flux and dispersion measure, we have had
to reduce the radio luminosity used in ACC by a substantial amount.

\section{Gamma-ray Emission Geometry}

For the geometry of the $\gamma$-ray beam, we have adapted the emission
from the slot gap described in Muslimov \& Harding (2003).  The slot
gap (Arons \& Scharlemann 1979) is a narrow region between two
conducting boundaries, the last open field line and the pair formation
front, extending from the neutron star surface up to the light cylinder.
Since the electric field is relatively small in the slot gap, primary
particles accelerate more slowly and pair cascades form at altitudes of
several stellar radii above the surface.  We model the radiation from
these pair cascades as having two components, curvature radiation of the
primary electrons and synchrotron radiation from the electron-positron
pairs.  We obtain the number of curvature, $N_{CR}$, and synchrotron,
$N_{SR}$, photons emitted per primary particle by integrating the
differential production rates over the available energy and over the
pulsar phase angle $\phi$. Since we have not included relativistic
effects such as aberration and time-of-flight delays in modeling the
$\gamma$-ray beam, the caustic peaks as found by Dyks \& Rudak (2003)
and in outer gap models (e.g. Romani \& Yadigaroglu 1995) do not appear
in our calculations.

\subsection{Curvature Radiation}

We assume that the curvature radiation takes place in the slot gap at
the last open field line (see Figure 1) and is integrated from a lower
$\gamma$-ray threshold, $E_\gamma$, of 100 MeV to the curvature
radiation critical energy, where $E_\gamma$ and $E_{CR}$, are in
$m_ec^2$ units and having an angular distribution represented by the
expression

\begin{eqnarray} \label{eq:curve} {{dN_{CR}}(>E_{\gamma},
\theta_{\gamma}) \over {d\Omega }}&=&{{3e^2} \over {2\pi \hbar c\sin
\theta }}\left( {{{\rho m_ec} \over \hbar }} \right)^{1/3}\left[
{E_{CR}^{1/3}-E_\gamma ^{1/3}} \right],\ {\rm where}\\ \nonumber
E_{CR}&=&{{3\gamma_o^3\hbar } \over {m_ec\rho }}\left[ {1+{{9\pi
r_e\gamma_o^3} \over {4cP}}\ln \left( {{r \over R}} \right)}
\right]^{-1},{\rm\ is\ the\ curvature\ photon\ critical\ energy\ and} \\
\nonumber \rho &=&{{r(1+\cos ^2\theta)^{3/2}} \over {3\sin \theta
(1+\cos \theta )}},{\rm\ is\ the\ radius\ of\ curvature\ with} \\
\nonumber r&=&{{cP} \over {2\pi }}\sin ^2\theta , \end{eqnarray}

\noindent where $r$ is the radial distance on the last open dipole field
line corresponding to magnetic colatitude, $\theta$, and
$\theta_\gamma={3\over 2}\theta$ is the photon emission angle, which is
tangent to this field line.   $R$ is the neutron star radius taken to be
$10^6$ cm, $r_e$ is the Compton wavelength of an electron and $\gamma_o$
is the initial Lorentz factor of the particle given by

\begin{eqnarray} \gamma_o&=&{{10^6B_{12}} \over {P^2}},{\rm\ for\
}B_{12}^{16/7}P^{-27/7}<1440,{\rm\ or}\\ \nonumber \gamma_o&=&4\times
10^7P^{1/14}B_{12}^{-1/7},{\rm\ for\ }B_{12}^{1/7}P^{-27/7}>3.0{\rm \
in\ regime\ I}\\ \nonumber \gamma_o&=&1.4\times 10^7P^{-1/4},{\rm \ for\
}B_{12}^{1/7}P^{-27/7}<3.0\ {\rm in\ regime\ II.} \label{eq:gamma}
\end{eqnarray}

\noindent where the first expression is for the case where the electric
field is not screened by electron-positron pairs (Harding et al. 2002)
and the second and third expressions are for cases where the electric
field is screened by pairs in the unsaturated (I) and saturated (II)
regimes (Zhang \& Harding 2000).  $P$ is the period in seconds and
$B_{12}$ is the magnetic field at the surface in units of $10^{12}$ G.
At a given inclination angle $\alpha$, the line of sight angle, $\zeta$,
and phase angle, $\phi$, define a polar angle, $\theta$ (through
equation [\ref{eq:theta}]), where the emission occurs tangent to the
last open field line at a radial distance, $r$, from the stellar center,
and the curvature emission rate per primary particle is given by
equation (\ref{eq:curve}) for $dN_{CR}/d\Omega$.

\subsection{Synchrotron Radiation}

The synchrotron radiation from cascade pairs takes place along the slot
gap, beginning at an altitude $R_{min}$, where the first pairs are
produced, and continuing out to $R_{SR}$, the maximum radius at which
pairs are produced (see Figure 1).  $R_{SR}$ is determined as the
altitude where the curvature radiation critical energy, $E_{CR}$, is
equal to the photon escape energy, $E_{esc}$ (i.e. the minimum energy of
photons capable of one-photon pair production).  The pair-escape
energy, $E_{esc}(r)$, in $m_ec^2$ units, is given approximately by (see
Zhang \& Harding 2000 or Harding 2001)

\begin{equation} E_{esc}(r) \approx 518\left( {{{rP} \over R}}
\right)^{1/2}\,\max \left( {1,{{0.1} \over {B'}}} \right).
\label{eq:esc} \end{equation}

The angles $\theta_{min}$ and $\theta_s$ are those corresponding to the
radii $R_{min}$ and $R_{SR}$ along the field line at the edge of the
slot gap,

\begin{eqnarray} \label{eq:theta_min} \theta_{\min}&=&\sin^{-1}\left(
{{{2\pi R_{\min }} \over {cP}}} \right),\\ \theta_s&=&\sin^{-1}\left(
{{{2\pi R_{SR}} \over {cP}}} \right). \end{eqnarray}

\noindent The corresponding photon emission angles are
$\theta_{\gamma,\min} \approx {3\over2}\theta_{\min}$ and
$\theta_{\gamma,s} \approx {3\over2}\theta_s$. The parameter,
$R_{min}/R$ is set to 3.5 in all of our simulations, fixing the
beginning of the emission zone to be 2.5 stellar radii above the surface
(Muslimov \& Harding 2003).  We assume that the electron-positron pairs
in the cascade have a spectrum $N_{\pm}(\gamma_p) = C_p \gamma_p^{-p}$,
extending from a minimum  $\gamma_{min} = E_{esc}(R_{\min})/2$ to a
maximum at $\gamma_{max} = E_{CR}/2$, where $E_{esc}(R_{\min})$ is the
photon pair-escape energy at radius $R_{\min}$.  

\noindent The integral photon spectrum above energy $E_{\gamma}$ of the
synchrotron radiation from the electron-positron pairs with spectral
index $p$, per primary particle, is

\begin{eqnarray} \label{eq:sync} {N_{SR}(> E_{\gamma})}&=&{{2 C_S} \over
{(1-p)}}\left[ {E_\gamma^{(1-p)/2}-E_{SR}^{(1-p)/2}} \right],{\rm\
where}\\ \nonumber \\ \nonumber  E_{SR}&=&{{3\gamma_{max}^2B'\sin \psi }
\over 2}. \\ \nonumber  \end{eqnarray}

\noindent $E_{SR}$ is the critical synchrotron energy, in $m_ec^2$
units, of pairs at their maximum energy $\gamma_{max}$ and $B'$ is the
local field strength in units of the critical field $4.4 \times 10^{13}$
G.  The pitch angle $\psi$ of the pairs is assumed to be that of the
parent photon direction with respect to the local field at the pair
production point, or

\begin{eqnarray} \sin \psi &=&{{0.1} \over {B'E_{CR}}},{\rm\ for\
}B'<0.1{\rm ,\ or}\\ \nonumber \sin \psi &=&{2 \over {E_{CR}}},{\rm\
for\ }B'>0.1.\\ \label{eq:psi} \nonumber \end{eqnarray}

\noindent The spectral index of the pairs is given by $p = 2\alpha_n -
1$, where the spectral index of the photons, $\alpha_n$, is determined
(Harding \& Daugherty 1999, Zhang \& Harding 2000) by the number
of generations, $n$, of the pair cascade,

\begin{eqnarray} \alpha _n&=&2-{{2-\alpha _1} \over {2^{n-1}}}, \\
\nonumber n&=&1+{{\ln (E_{esc}/E_o)} \over {\ln (\kappa )}}\\
\label{eq:an} \end{eqnarray}

\noindent where $\kappa = 3/64$ and $\alpha_1=5/3$ for curvature
radiation with losses. The cascade generation number is determined by
$E_o$, which is the curvature radiation critical energy at the initial
colatitude angle $\theta_{min}$ of the radiation zone.

\noindent We normalize the pair spectrum to the total cascade
multiplicity

\begin{equation} N_{\pm} \sim {E_0\over 20 E_{esc}(R_{\min})}
\end{equation}

\noindent so that the normalization factor $C_S$ for the synchrotron
photon spectrum in equation (\ref{eq:sync}) is given by

\begin{equation} C_S={{4\pi } \over {3.3\sqrt 3}} {2^{1-p}(1-p)\over
(3p-1)}N_{\pm}\,[E_0^{(1-p)}-E_{esc}(R_{\min})^{(1-p)}]. \label{eq:cs}
\end{equation}

\noindent The emission from the high altitude (~2-4 stellar radii)
cascades from the slot gap along the last open field line forms a broad,
hollow-cone beam.  The parameter representing the longitudinal thickness
of the slot gap is expressed, in units of the polar cap half-angle
$\theta_{\min}$, as (Muslimov \& Harding 2003)

\begin{eqnarray} \Delta \xi &=&4PB_{12}^{-4/7}\ {\rm for}\
PB_{12}^{-4/7}<0.075{\rm\ or}\\ \nonumber \Delta \xi &=&0.3\ {\rm for}\
PB_{12}^{-4/7}>0.075,\\ \label{eq:xi} \nonumber  \end{eqnarray}

\noindent where $B_{12}$ is in units of $10^{12}$ G.  The
acceleration-cascade simulations indicate that the width of the slot
widens as the pulsar ages and saturates at a value of approximately 0.3.
As seen in Figure 1, the interior and exterior polar angles of the
radiation from the slot gap at $R_{\min}$ are described by the following
expression

\begin{eqnarray} \theta_{\min }^{SG}&\sim& \theta_{\gamma,\min}
(1-\Delta \xi ),{\rm\ and}\\ \nonumber \theta_{\max }^{SG}&\sim&
\theta_{\gamma,\min}\\ \label{eq:thetasg} \nonumber  \end{eqnarray}

\noindent We take the average opening angle of the cascade radiation
from the slot gap between $r = R_{\min}$ and $r = R_{SR}$ as
$\theta_{SG} = (\theta_{\min }^{SG} + \theta_{\gamma,s})/2$. We
approximate the angular distribution of the synchrotron radiation
component of the entire cascade between $\theta_{min}$ and $\theta_s$ as
a hollow beam with a conal Gaussian of width equal to

\begin{equation} w_\gamma =\theta _{_{\max }}^{SG}-\theta _{_{\min
}}^{SG}, \label{eq:wsg} \end{equation}

\noindent which is the full width at $1/e$ of the maximum. The integral
photon spectrum above energy $E_{\gamma}$ of the synchrotron radiation,
per primary particle at a given polar angle is then given by

\begin{equation} {{dN_{SR}(> E_{\gamma},\theta_{\gamma})} \over
{d\Omega}} = {N_{SR}(> E_{\gamma}) \over
\theta_{SG}\omega_{\gamma}\pi^{3/2}}
\exp\left[-{(\theta_{\gamma}-\theta_{SG})^2\over
\omega_{\gamma}^2}\right] \end{equation}

\noindent The current of primary electrons in the slot gap that results
in curvature and synchrotron radiation is limited to a fraction of the
Goldreich-Julian current, $\dot n_{GJ}$, in the following manner

\begin{eqnarray} \dot n_{slot\ gap}&=&(1-(1-\Delta \xi )^2)\dot
n_{GJ},{\rm\ where}\\ \nonumber \dot n_{GJ}&=&{{5.7\times
10^{31}B_{12}^{\scriptstyle {}\hfill\atop \scriptstyle {}\hfill}} \over
{P^2}}.\\ \label{eq:rhosg} \end{eqnarray}

\noindent This current multiplies the integral of he curvature and
synchrotron emission per primary particle to give the total slot gap
emission beam.

The total flux due to both curvature and synchrotron radiation are
calculated for a given phase angle, $\phi$, which is related to
$\theta_{\gamma}$ through equation (\ref{eq:theta}), for a pulse profile
with 500 bins of phase angle from $-\pi$ to $\pi$.  The average of the
profile is obtained and compared to the appropriate instrumental flux
threshold.  If the average flux is above the threshold, the $\gamma$-ray
pulsar is ÒdetectedÓ. This condition is tested independently of the
radio flux and appropriate radio survey threshold, allowing us to
designate the ÒdetectedÓ $\gamma$-ray pulsar as radio-loud or
radio-quiet.

\section{Monte Carlo Simulations}

We discuss here some of the important changes that have been made to our Monte 
Carlo simulation code from our previous work in Gonthier et al. (2002).
While we believe that is important to place the neutron stars at birth
in spiral arms, we have not yet included the spiral arm structure into
our simulations.  As in Gonthier et al. (2002), we distributed pulsars
at birth in the Galactic disk according to the prescription of
Paczy\'{n}ski (1990). In a cylindrical coordinate system, the azimuthal
angle, $\phi$, is randomly chosen between 0 and $2\pi$.  The $z$
distribution varies exponentially with distance from the plane, while
the radial distribution peaks at 4.5 kpc and decreases exponentially
from the center of the Galaxy.  Given the initial position and velocity,
the trajectory of each neutron star is evolved in the Galactic potential
to the present.

\subsection{Comparison group of pulsars in the ATNF catalog}

In order to have a comparison group to normalize our simulation, we have
selected pulsars from the Australian Telescope National Facility
(ATNF) \footnote{ATNF catalog is available at:
http://www.atnf.csiro.au/research/pulsar/psrcat/.}. We chose pulsars
within the Galaxy with periods greater than 30 ms and with positive
period derivatives to obtain a comparison group of 978 pulsars detected
by these nine surveys.  Selecting pulsars with periods greater than 30
ms insures that we have left out of our group most of the millisecond
pulsars that have been recycled in binary systems. We are not currently
simulating this class of pulsars since their evolution is more
complicated. We have also not included the anomalous X-ray pulsars, the
soft-gamma-ray repeaters, or pulsars in globular clusters in our
comparison group.  We run the Monte Carlo simulation until the code
ÒdetectsÓ the same total number of radio pulsars as have been observed
with the group of surveys. With this normalization a neutron star birth
rate is predicted as well as the number of $\gamma$-ray pulsars detected
by various instruments. However, in order to obtain smoother simulated
distributions, we run the code for ten times the number of pulsars
detected by the radio surveys and then normalize accordingly.

\subsection{Flux sensitivity of the Parkes multi-beam pulsar survey}

We have included the eight radio surveys described in Gonthier et al.
2002 along with the new PMBPS having an angular coverage of  $\mid
b\mid<5^\circ$ and $\ell=260^\circ$ to $\ell=50^\circ$ with an assumed
geometric efficiency of 100\%.  In Gonthier et al. 2002, we calculated
the minimum radio thresholds, $S_{min}$, for the selected group of radio
surveys using the Dewey et al. (1985) formula.  We attempted to use the
same formula for the PMBPS using the parameters indicated in Manchester
et al. (2001).  However, we found that an additional factor of $\sim 2$
multiplying the limiting sensitivity is required to reproduce the
$S_{min}$ curves in Figure 2 of Manchester et al. (2001) as shown here
in Figure 2.  The Dewey formula under predicts the $S_{min}$, and a more
realistic treatment of narrow pulse widths (smaller duty cycles) in the
Fourier search is not as optimistic as the Dewey formula (F. Crawford
private communication).  As a result, we chose to evaluate the $S_{min}$
for the PMBPS using an IDL code (Crawford private communication) that we
translated into C++ to incorporate into our Monte Carlo code that
is called event-by-event.  This routine was used to create Figure 2 in
Manchester et al. (2001) and reproduced here in Figure 2, along with the
$S_{min}$ curves predicted by the Dewey formula, with the extra factor
of 2, for the indicated DMs.  In our simulations, we have then scaled
the limiting sensitivities as discussed in Manchester et al. (2001).

\subsection{Gamma-ray Thresholds}

We simulate the $\gamma$-ray pulsars ÒdetectedÓ by EGRET, AGILE and
GLAST.   If the simulated $\gamma$-ray flux, obtained from the average
flux in the pulse profile, is above a detector threshold, the pulsar is
said to be a $\gamma$-ray pulsar detected by the corresponding
instrument.  We have included an all-sky sensitivity for both EGRET (I.
Grenier, private communication) and AGILE (A. Pellizzoni, private
communication) and are shown in Figure 3a and 3b.  For AGILE, we have
three all-sky sensitivity maps representing the best, nominal and worst
case scenarios.  The one portrayed in Figure 3b is for the nominal case.
For GLAST, we have used the following thresholds: in-plane ($\mid
b\mid<10^\circ$) $5\times 10^{-9}\ \rm photons/(cm^2\cdot s$),
out-of-plane ($\mid b\mid\geq 10^\circ$) $2\times 10^{-9}\ \rm
photons/(cm^2\cdot s$) (D. Thompson private communication), and for
pulsed emission $5\times 10^{-8}\ \rm photons/(cm^2\cdot s$) (S. Ritz
private communication, McLaughlin \& Cordes 2000).  The above threshold
for pulsed emission detection in a blind periodicity search is based on
techniques used in periodicity searches of EGRET data (Mattox et al.
1996, Chandler et al. 2001).

\subsection{New distance model}

We have incorporated the new electron density model of the Galaxy from
Cordes \& Lazio (2003) by calling the FORTRAN subroutines from within
our code to calculate the dispersion measure (DM) of the simulated
pulsar. The DM leads to a smearing of the pulse, affecting the flux
threshold for radio detection.  For comparison, we have recalculated the
distance of the pulsars in the ATNF catalogue from the measured DM and
the pulsarÕs location using the new distance model.  In Figure 4, we
show the histogram for the logarithm of the absolute value of the
difference between the distance obtained from the new distance model and
the old distance model for our selected group of 978 pulsars. The
distances obtained with the new distance model are about 20\% smaller
than those obtained with the old distance model.  For pulsars in the
catalogue whose Òbest estimateÓ distance is different than the one
obtained using the old distance model, we have assumed that the distance
was established by other methods and, therefore, is more reliable.

\subsection{Initial period distributions}

Recently various observations of young supernova remnants have been able
to measure the speed of the expansion shell and the period and period
derivative of the pulsar, thereby, determining the initial period of the
pulsars.  For example, X-ray Pulsars PSR J1811-1925 and PSR J0205+6559
(Gavriil et al. 2003) have been associated with the supernova remnants
G11.2-0.3 and 3C 58, respectively, and may suggest that these pulsars
where born with a period of $\sim 65$ ms.  In contrast to our previous
study of Gonthier et al. (2002) that used a constant birth period of 30
ms, we studied Gaussian and flat initial spin distributions to describe
the initial period.  We found that the overall population statistics are
not very sensitive to the initial spin distribution and only affects the
short-period population of pulsars in the $\dot{P}-P$ diagram.  While
significant progress is being made in deducing the initial period of
pulsars, the shape of the distribution is not well defined at the
present.  We have concluded that a flat distribution from 0 to 150 ms
accommodates the observations and have used this distribution in this
study.

\subsection{Decay of the Magnetic Field}

We continue to be steered in the direction of incorporating the decay of
the magnetic field in order to achieve better comparisons. Originally,
we included eight radio surveys in Gonthier et al. (2002) with 445
detected radio pulsars to compare to our simulated results. We used one
Gaussian to describe the primary magnetic field distribution with a
single decay constant.  The PMBPS (Manchester et al. 2001) has
discovered many more pulsars, many of which are young, distant pulsars
with high radio luminosities.  The current pulsar catalog now has 1412
radio pulsars (http://www.atnf.csiro.au/people/pulsar/catalogue).  With
the PMBPS, we have a selected group of 978 detected radio pulsars with
many more high field pulsars.  A single Gaussian would result in too
many low field pulsars.  In order to simulate these high field pulsars,
we found it necessary to use two Gaussian distributions to skew the
distribution towards high fields.

The pulsarÕs surface magnetic field distribution at birth is represented
by the sum of two log-normal Gaussian distributions expressed as

\begin{equation} \rho_B=\sum\limits_{i=1}^2 {A_ie^{-(\log B-\log
B_i)^2/\sigma _i^2}}, \label{eq:rhob} \end{equation}

\noindent where the parameters are indicated in Table 1.

\begin{center} \begin{tabular}{cccc} \multicolumn{4}{c}{Table 1} \\
\hline \multicolumn{4}{c}{Parameters for Initial}\\
\multicolumn{4}{c}{Magnetic Field Distribution}\\ \hline $i$ &$A_i$  &
$\log B_i$	& $\sigma_i$\\ \hline 1 &0.6	& 12.75 &0.4 \\ 2
&0.3	& 13.0  &0.7 \\ \hline \end{tabular} \end{center}

\noindent While there are two Gaussians describing the initial field
distribution of the pulsars at birth, the second Gaussian with a higher
mean field merely skews the distribution towards higher magnetic
fields and does not necessarily suggest two groups of pulsars with
different field characteristics.  Using a single broader Gaussian would
result in too many lower field pulsars.

The birth rate is assumed to be constant during the history of the
Galaxy (at least back to $10^9$ years in the past); therefore, we
randomly select the age of the pulsar from the present to $10^9$ years
in the past. We assume a dipole spin down with a decaying magnetic field
having a time constant, $\tau_D$, following Gonthier et al. (2002).  In
Figure 5, we present the period derivative versus the period of
simulated pulsars with the indicated time constants from $10^8$ to
$5\times 10^5$ years.  Indicated in the figure are lines of constant
field (calculated according to Shapiro \& Teukolsky, 1983) and pulsar
age assuming dipole spin-down of a constant field as well as the
curvature radiation (CR) and nonresonant inverse Compton scattering
(NRICS) death lines (Harding et al. 2002).  Without field decay or a
large decay constant like $10^8$ years, pulsars will move from their
short periods to longer periods along constant or nearly constant field
lines and pile up near the NRICS line.

In order to populate, without field decay, the diagram in a region of
small period derivatives ($5\times10^{-18}$) and medium periods (0.5 s),
many more short period pulsars would populate the lower left region of
the diagram where none are observed (in Figure 8).  Decreasing the decay
constant, produces the upside-down pear shaped distribution seen in the
distribution of detected pulsars and populates the high field region
above $5\times 10^{13}$ G.  Unless one can alter the period and period
derivative dependence of the radio luminosity significantly in a manner
with more than just a simple power law, we find that field decay is
required to reproduce the distribution.  In subsequent simulations, we
have adopted a value of 2.8 Myr for the decay constant.

\subsection{Supernova kick velocity distribution}

A number of studies disagree on the initial 3-D velocity distribution of
neutron stars at their birth, possibly the result of an asymmetric
supernova kick, typically described by a Maxwellian distribution.
Lorimer, Bailes \& Harrison (1997) obtained a velocity distribution with
a mean velocity of $\sim$480 km/s similar to a previous study obtaining
a mean of $\sim$450 km/s (Lyne \& Lorimer 1994), yet significantly
larger than most previous studies of pulsar statistics that required
space velocities of $\sim$150 km/s.  Hansen \& Phinney (1997) concluded
that a mean velocity $\sim$250-300 km/s best described their study.
Though Hartman et al. (1997) did not use a Maxwellian distribution, they
obtained a distribution with a mean velocity of 380 km/s.  Gonthier et
al. (2002) also did not use a Maxwellian distribution and found a
distribution with a mean velocity of 170 km/s.  It is clear that the
velocity distribution that one obtains depends heavily on the many other
assumptions that go into the model, such as the radio luminosity and
radio beam geometry. The brighter the radio pulsars, the greater the
distance at which they are detected, resulting in a broader distribution
of distance, $z$, from the plane of the Galaxy, requiring a smaller mean
velocity to improve the agreement with $z$ distribution of detected
pulsars.

In this study we have adopted the luminosity model of ACC, and so we
must also adopt their kick velocity model.  We chose to follow their
two-component velocity distribution, which is Maxwellian in velocity
with characteristic widths of 90 and 500 km/s and given by the equation
(1) in ACC. In the ACC model, the two-component velocity model was
preferred over the single component model.  The fraction of the neutron
stars with a width of 90 km/s is 40\% and with a width of 500 km/s is
60\%, leading to an average velocity of $\sim$540 km/s.

In Figure 6, we show the $z$ distribution above the Galactic disk for
the detected pulsars (shaded histogram) and for the simulated pulsars
(regular histogram).  Under the assumptions of the model, the predicted
distribution is a little wider that the one for detected pulsars, having
scale heights of 152 pc and 182 pc, respectively.  We realize that
many assumptions in our model are interrelated and decreasing the
overall radio luminosity of the ACC model leads to a difference in the
$z$ distribution as pulsars are radio dimmer and must be closer to be
ÒdetectedÓ.  We have chosen to keep the velocity model and overall radio
luminosity model of ACC making as few necessary changes to the ACC model
as needed.

\subsection{Reduction of the radio luminosity}

Using the radio luminosity of ACC, we find that the simulated radio
pulsars are too bright, with too many distant pulsars being
detected and
predicting a neutron star birth rate of 0.11 per century, with no
$\gamma$-ray pulsars predicted to be detected by EGRET.  In ACC
all the pulsar surveys used in the study were at frequency near 400 MHz.
In the set of surveys chosen for this study there are two groups with
one having frequencies near 400 MHz and the other group with frequencies
near 1400 MHz.  Since the PMBPS $S_{min}$ is best accounted for in our
simulation and this survey at 1374 MHz detected most of the pulsars
observed in the Jodrell Bank 2 survey at 1400 MHz and the Parkes 1
survey at 1520 MHz, we selected only the PMBPS pulsars to represent the
Òhigh frequencyÓ (HF) surveys, while the other selected surveys in our
study near 400 MHz represent the Òlow frequencyÓ (LF) surveys.  Focusing
primarily on the distributions of the pulsar distance, DM and flux at
400 MHz and 1400MHz, we first set the spectral indices (preserving
0.5 between core and cone indices) to give the same birth rate for these
two frequency groups, then we set the over all luminosity to given a
reasonable birth rate of $\sim 1.5$ neutron stars per century.  We obtain the
same birth rate for each group with spectral indices of
$\alpha_{core}=-2.1$ and $\alpha_{cone}=-1.6$.  These spectral indices
describe the primary spectra of the radio pulsars before the selection
effects of the characteristics of the chosen radio surveys.

In our simulation, we calculate the
fluxes at the frequency of each of the surveys in our selected group as
well as at 400 and 1400 MHz by averaging the pulse profile for a given
random viewing geometry for each pulsars.  From the calculated S400 and
S1400, we obtain a spectral index for each pulsar.  In the simulation
with all the surveys, we simulated 9780 (ten times the number in our
select group of surveys to improve statistics) radio pulsars as
ÒdetectedÓ by these surveys and find an average spectral index of -1.8
and a standard deviation of 0.2.  Lorimer et al. (1995) measured
spectral indices of 280 pulsars by measuring fluxes at radio frequencies
between 408 and 1606 MHz with the distribution having a dependence on
the characteristic age of the pulsar with a dependence,
$\alpha=-1.7+0.2\log\left( \dot{P}/ P \right)$, with a standard
deviation of 0.6 with respect to this dependence. Maron et al. (2000)
extended to lower and higher frequencies the study of Lorimer et al.
(1995) and obtained an average spectral index of -1.8 with a standard
deviation of 0.2.

To show the overall effect of the reduction of the radio
luminosity used in ACC, we show in Figure 7 the comparisons of the
distributions for the pulsar distance, flux at 1400 MHz and DM for 620
pulsars detected with only the PMBPS and an equal number simulated
pulsars for this survey alone. Given that we are calculating the
$S_{min}$ for the PMBPS according to the formulation used in Manchester
et al. (2001), we believe that the $S_{min}$ for this survey is the best
described in the group of our selected surveys, and with all pulsars
detected at $\sim$1400 MHz sky temperature effects are minimized,
reducing further uncertainties. Indicated in Figure 7 are the factors,
$f_{red}$, used to reduce the radio luminosity of the ACC model and the
resulting distributions.  Since the ACC
model studied pulsar surveys at 400 MHz, the factor $f_{red}$ represents
a reduction of the ACC 400 MHz luminosity by this factor.  As the radio
luminosity is reduced, the comparison of the distances and DMs improves,
but disagreement increases between the distribution of the simulated and
detected radio flux at 1400 MHz. There are significantly more pulsars
simulated with lower radio fluxes than those detected by this survey,
suggesting that perhaps certain aspects of the emission geometry are
still not adequately described. We find that the shape of the flux
distribution is not very sensitive to the features of the conal
geometry, such as the radius and width.  We also wanted to predict a
reasonable  neutron star birth rate per century, which varies from 
0.6 at $f_{red}=20$ to 1.6 at $f_{red}=80$.  We chose to reduce the
radio luminosity by a factor of $f_{red}=60$ in subsequent simulations
in the following figures compromising between good agreement of the
distances and DMs and less desirable agreement of the fluxes. These
factors predict a birth rate of 1.38 neutron stars per century for the
case of all nine radio surveys.

\section{Results}

In Figure 8, we present the distribution of our select group of 978
detected pulsars in the ATNF pulsar catalog (8a) and the same number of
simulated pulsars (8b) as a function of period derivative and period. The
solid lines represent constant dipole magnetic field of $10^{11-14}$ G,
and the dashed lines correspond to the curvature and nonresonant inverse
Compton scattering death lines.  As dotted curves, we show paths in the
diagram for 4 pulsars assuming a field-decay constant of 2.8 Myr, all
with ages of $10^7$ years, and with the indicated initial magnetic
fields in $B_{12}$ units.  We show the age lines assuming field decay as
dot-dashed lines. Due to field decay, these lines are very different
from the characteristic age lines, with the oldest pulsars being 10 Myr
rather than 1 Gyr assuming no field decay.   As indicated in Table 1, we
chose to represent the initial magnetic field distribution with two
Gaussians. The initial magnetic fields of the four evolutionary paths
portrayed in Figure 8 were chosen to represent the means of the two
Gaussians, the higher mean plus the high field component's width and the
lower mean minus the lower field component's width. As a result, most of
the simulated pulsars will lie within the lowest and highest paths.  Due
to a 2.8 Myr decay constant, the knee-like, curved portions in these
paths begins before 1 Myr and become vertical after a few Myr.  The
paucity of pulsars in the two regions indicated with the circle and LB
and HB for low and high field (Figure 8a) can be explained in terms of
decay of the surface magnetic field on a time scale of 2.8 Myr.  While
it may be true that one can tailor a radio luminosity law to account for
the observed distribution, field decay leads naturally to the
funnel-shaped distribution of the detected pulsars.

In Figure 9, we present the comparisons between detected and simulated
statistics for the indicated parameters.  The shaded histograms
represent the distributions of the detected pulsars while the regular
histograms correspond to the model simulations. The model simulation
over-predicts the number of pulsars with short periods, large
period derivatives and larger distances.  Improved comparisons can be
obtained by making the radio luminosity not as strongly dependent on the
period by decreasing the exponent from -1.3 to -0.9 and that of the
period derivative from 0.4 to 0.3.  The main discrepancy lies in the
comparisons of the distance distributions and flux distributions (Figure
5).  The agreement in the flux distributions for only the PMBPS, shown
back in Figure 5, is better with a reduction factor of $f_{red}=20$
making the pulsars more radio bright. However, the agreement between the
distance distributions is significantly worse and the birth rate is too
low.  Perhaps the inability to find agreement in the radio flux and
distance distribution may stem from our underlying assumption that radio
pulsars are standard candles or may be indicating that spiral arm
structure may have to be used for the birth location of neutron stars in
order to improve the agreement between the simulated and observed
distributions.

In Figure 10, we show the distributions in Galactic coordinates as
Aitoff projections for 978 pulsars detected (10a) and simulated (10b).
The strong contribution to the Galactic disk is due primarily to the
PMBPS, adding nearly half of the total number of pulsars that are unique
to this survey.

We also simulate the $\gamma$-ray pulsar detections by EGRET, AGILE and
GLAST, using the assumptions discussed earlier, and independently
simulate detection of radio pulsars.  However, the $\gamma$-ray pulsars
are flagged as radio-loud if their fluxes are higher than the minimum
sensitivities of the select group of surveys; otherwise they are radio-quiet.
Therefore, we can predict the number of radio-loud and radio-quiet
$\gamma$-ray pulsars detected as point sources by each of the three
instruments.  For known radio pulsars, with measured periods and period
derivatives, the $\gamma$-ray instrument can detect them as point
sources and obtain a pulsed detection through reliable epoch folding.
GLAST will have the sensitivity and ability to perform blind period
searches and detect pulsation without a radio ephemeris. Table 2
indicates the simulated pulsar statistics for the radio-quiet and
radio-loud $\gamma$-ray pulsars as detected by the three instruments.

\begin{center} \begin{tabular}{c|c|c|c|c} \multicolumn{5}{c}{Table 2} \\
\hline \multicolumn{5}{c}{Simulated Pulsar Statistics }\\ \hline\hline
&\multicolumn{2}{c|}{All Nine Surveys}&\multicolumn{2}{c}{All Surveys
Excluding PMBPS}\\ &\multicolumn{2}{c|}{(978 radio
pulsars)}&\multicolumn{2}{c}{(546 radio pulsars)}\\ \hline Birth Rate &
\multicolumn{2}{c|}{1.38} & \multicolumn{2}{c}{1.46} \\ \hline &Radio
Quiet & radio-loud & radio-quiet & radio-loud \\ \hline EGRET & 7 & 19 &
10 & 15 \\ AGILE & 13 & 37 & 19 & 36\\ GLAST & 276 & 344 & 436 & 209\\
\hline \end{tabular} \end{center}

To improve the statistics, we run the simulation until the number of
radio pulsars ÒdetectedÓ by the chosen set of radio surveys is equal to
ten times the number of actual detected radio pulsars by those surveys.
We then renormalize by dividing by a factor of ten.  EGRET observed all
of the radio pulsars that were detected by all the radio surveys
excluding the PMBPS, out of which EGRET detected 8 $\gamma$-ray pulsars
including Vela, the Crab, B1951+32, B1706-44, B1055-52, B0656+14,
J1048-5832, and the radio-quiet $\gamma$-ray pulsar, Geminga  plus a
couple of other candidate pulsars bring the total to perhaps 12 (Kanbach
2003).  Excluding the PMBPS, our simulation predicts that EGRET should
have seen 15 radio-loud and 10 radio-quiet $\gamma$-ray
pulsars and a neutron star birth rate of 1.46 per century. With
all nine surveys, the predicted birth rate of 1.36 is 5\%
smaller and, therefore, the total number of $\gamma$-ray pulsars has
also dropped by 5\% (GLAST) within statistics.  The EGRET Third
catalogue (Hartman et al. 1999) contains 170 unidentified point sources,
some of which are expected to be pulsars.  Sensitive searches performed
with Chandra and the Parkes multi-beam telescopes have resulted in a few
new pulsars within the error boxes of the unidentified EGRET sources
(Halpern et al 2001; D'Amico et al. 2001).  Correlating the positions of
the radio pulsars detected in the PMBPS with the EGRET unidentified
sources, Torres et al. (2001) found 14 positional coincidences.  With
the nearly completed PMBPS, Kramer et al. (2003) found about 38
positional coincidences, and they determined that $19\pm 6$ are
statistically likely to be real associations.  So it would seem then
that adding the PMBPS should convert radio-quiet $\gamma$-ray pulsars
into radio-loud $\gamma$-ray pulsars as detected by EGRET.  This is
clearly the case for GLAST and there is a significant conversion of
radio-quiet to radio-loud pulsars for AGILE and EGRET when the PMBPS is
added. Since our present simulation over-predicts the number of short period,
energetic pulsars, mostly detected by the eight surveys (without PMBPS)
prior to EGRET, our results are over-predicting the number of radio-loud
pulsars detected by EGRET from these surveys.  However, the number of
radio-loud pulsars predicted for all surveys including PMBPS is
consistent with the number of plausible coincidences found by Kramer et
al. (2003).

In Figure 11, we present the positions in the $\dot{P}-P$ diagram of the
known $\gamma$-ray pulsars detected by EGRET (10a) and of those
simulated for EGRET (10b), AGILE (10c) and GLAST (10d) where radio-loud
$\gamma$-ray pulsars are shown as solid circles and radio-quiet
$\gamma$-ray pulsars are shown as crosses.  Younger pulsars have higher
$\gamma$-ray luminosities that decrease as they approach the curvature
death line where curvature radiation $\gamma$-rays can no longer produce
electron-positron pairs.  The $\gamma$-ray luminosity decreases
significantly for older pulsars below the curvature radiation death line
(Harding et al. 2002), where the main mechanism of pair production is
via inverse Compton scattering of the thermal soft X-rays from the
stellar surface.  Pulsars below the nonresonant Compton scattering death
line are unable to produce pairs and become radio-quiet.  Our
simulations predict that GLAST will detect 276 radio-quiet and
344 radio-loud $\gamma$-ray pulsars; a significant improvement
over EGRET.  With GLAST ability to perform blind period searches, we
predict that out of the 276 radio-quiet $\gamma$-ray pulsars, 17 pulsars will be detected as pulsed sources. AGILE, scheduled to
launch before GLAST, should detect 13 radio-quiet and 37
radio-loud $\gamma$-ray pulsars using the nominal sensitivity map. The
best and worse case-maps predict 24 and 9 radio-quiet and
64 and 27 radio-loud $\gamma$-ray pulsars, respectively. As
all-sky sensitivity and threshold maps are improved, these numbers will
vary somewhat.

We present in Figure 12 the Aitoff projections of the $\gamma$-ray
pulsars observed by EGRET and those simulated for EGRET, AGILE and
GLAST.  Most of the pulsars detected by GLAST will be young pulsars with
ages $\sim 10^5$ years and within 500 pc of the Galactic disk.  The
asymmetric distribution of radio-loud and radio-quiet $\gamma$-ray
pulsars is a result of the PMBPS detecting radio pulsars from $\ell =
260^\circ$ to $\ell = 50^\circ$, mostly on the right side of the figure.

In order to further explore the effect of the parameterization of the
beam geometries of the radio and $\gamma$-ray emission, one has to study
the pulse profiles and the correlation between the radio profile and the
$\gamma$-ray profile.  In Figure 13, we present a select pair of
examples of simulated radio and $\gamma$-ray profiles for two radio-loud
pulsars ``detected'' by EGRET with periods of 528 ms (13a) and 41 ms
(13b).  The two pulsars also have similar impact angles of $4.1^\circ$
(13a) and $-5.5^\circ$ (13b). The radio profiles presented here are for
400 MHz and the $\gamma$-ray profiles are for greater than 100 MeV.  The
ACC model predicts ratios of the radio core-to-cone flux of 13 (Figure
13a) and 163 (Figure 13b), suggesting that the conal contribution to the
flux is minor.  However, the viewing geometry is important also.  The pulse profile in
 the line-of-sight
intersects the radio cone and the outer portion of the radio core,
displaying two radio peaks, but only one curvature radiation
$\gamma$-ray peak.  Due to the large impact angle, the radio core does
not significantly contribute to the overall profile as this is a fairly
long period pulsar.  The other pulsar in Figure 13b has a similar impact
parameter but with a shorter period, therefore the radio core clearly
dominates the profile and the two conal peaks from the $\gamma$-ray
pulse are manifested.  Since in this model the $\gamma$-ray emission
originates within 2.5 stellar radii of the surface and the radio core
emission is  believed to come from similar altitudes above the stellar
surface, there would be little aberration or time delay between them and
they should appear correlated with the radio core peak being in phase
with the single $\gamma$-ray peak, as in Figure 13a, or being in between
the two $\gamma$-ray peaks, as in Figure 13b.  On the other hand the
radio cone beam is believed to arise from a higher altitude region and
the effects of aberration and time delay, not included in our model,
might shift the peaks to earlier phase relative to the $\gamma$-ray or
radio core in the pulse profile. This effect is actually observed in a
number of triple radio profiles, where the core component lags the
center of the cone (Gangadhara \& Gupta 2001).  In the ACC model with
core dominated short-period pulsars, we find that 54\% of the EGRET
radio-loud $\gamma$-ray pulsars have two $\gamma$-ray peaks in the pulse
profile and are all core dominated, exhibiting a single (core) radio
peak in the profile.  The other 46\% have a single $\gamma$-ray peak
with a variety of radio profiles, including a single core peak, core and
conal peaks, and two conal peaks.  However, the fact that all two-peaked
$\gamma$-ray profiles are core dominated is contrary to the few EGRET
$\gamma$-ray pulsars, typically consisting of two $\gamma$-ray peaks
with a single radio peak leading the $\gamma$-ray peaks (Thompson et al.
1997).  This significant discrepancy between the observed and simulated
radio and $\gamma$-ray profile correlations either questions the
parameterization of the core to cone radio flux in the ACC model
suggesting that these short period pulsars are different, having their
profiles dominated by cone emission, or questions the polar cap
$\gamma$-ray emission model we have used.

By contrast, several recently discovered, young X-ray pulsars show
single broad peaks in their profiles and also single radio peaks in
phase with the X-ray peaks (Lorimer 2003 and references therein). They
also have very low radio flux and luminosity. Harding et al. (2003) have
shown that the radio characteristics can be understood in the ACC model
if we are viewing the edge of the cone beams at large impact parameter,
so that the cone beam appears single and the core component is not seen.
In the polar cap model emission geometry, the high-energy emission peak
would be in phase with the single radio peak, as observed. A number of
these are coincident with EGRET sources and AGILE may measure the
$\gamma$-ray profiles.

\section{Discussion} We have included radio and $\gamma$-ray beam
geometries into our Monte Carlo code that simulates population
statistics for radio and $\gamma$-ray pulsars, making predictions for
the number of radio-quiet and radio-loud $\gamma$-ray pulsars detected
by the instruments: EGRET, AGILE and GLAST.  The radio beam geometry is
tailored after the phenomenological model of ACC with slight
modifications to include the radius-to-frequency mapping of Mitra \&
Deshpande (1999).  In the ACC model, radio pulsars are assumed to be
standard candles with their radio luminosity described by a simple power
law in period and period derivative.  The $\gamma$-ray beam geometry has
been derived from the theoretical work of Muslimov \& Harding (2003)
describing the emission from the slot gap in the polar cap model.  These
enhancements are significant improvements over our previous studies of
Gonthier et al. (2002).  We have added the PMBPS to our select group of
radio surveys, we have used the new distance model of Cordes \& Lazio
(2003), we have added all-sky threshold maps for EGRET (Grenier private
communication) and AGILE (Pellizzoni private communication), and we have
used realistic $S_{min}$ thresholds for the PMBPS (Crawford private
communication).  Neutron stars are assumed to be born with a constant
birth rate, with a Galactic distribution described by Paczy\'{n}ski
(1990) with a flat distributions in their initial period and a
Gaussian distribution in their magnetic field, and their trajectories
are evolved to the present within the Galactic potential (Paczy\'{n}ski
1990).  We find better agreement when the magnetic field is assumed to
decay exponentially with a decay constant of 2.8 Myr, which is somewhat
shorter than the decay constant in Gonthier et al. (2002) that
assumed an entirely different beam geometry and radio and $\gamma$-ray
luminosity models.

In order to obtain agreement between observed and simulated radio pulsar
distance distributions, a reasonable birth rate of 1.45 neutron stars
per century and a reasonable number of $\gamma$-ray pulsars observed by
EGRET, we find it necessary to significantly reduce by a factor 60 the
overall radio luminosity of ACC model.  The major problem with our radio
emission model is that we are not able to simultaneously fit the
distance distribution and the radio flux distribution as seen in Figure
5.  We found that slight adjustments to the radio beam geometry do not
affect these distributions.  This inadequacy perhaps challenges some
major assumptions of the model such as that radio pulsars are standard
candles or that short period pulsars are core dominated.  There is a
lack of theoretical insight into the mechanism for the radio luminosity.
Randomizing the radio luminosity about the expected value does indeed
achieve some improvement in the flux distribution.  The disagreement
between the inferred and the simulated distance distributions may be a
result of assuming a smooth distribution of neutron stars in the
Galactic plane, rather than taking into account the spiral structure of
the Galaxy.  It is very apparent that most pulsars are detected in
spiral arm regions.

Our simulation is normalized to the number of radio pulsars observed by
the selected group of surveys and, therefore, a neutron star birth rate
is predicted, as well as the number of $\gamma$-ray pulsars observed by
each instrument as shown in Table 2.  We expected by adding the PMBPS
that the ratio of radio-quiet $\gamma$-ray pulsars to radio-loud
$\gamma$-ray pulsars would decrease for EGRET given the positional
coincidences found between the Parkes multi-beam radio pulsars within
EGRET error boxes.  Recently Kramer et al. (2003) estimated about $19\pm
6$ associations, which suggests about 20 radio-quiet $\gamma$-ray
pulsars detected by EGRET that become radio-loud when the PMBPS is
included. The simulated ratio of radio-loud to radio-quiet
$\gamma$-ray pulsars decreases by 45\% from 2.7 (with PMBPS) to 1.5
(without PMBPS) for EGRET.  For AGILE there is also a similar decrease
of 34\% from 2.9 (with PMBPS) to 1.9 (without PMBPS), but very evident
for GLAST as the ratio decreases significantly by 62\% from 1.25 (with
PMBPS) to 0.5 (without PMBPS).

Our model assumes that pulsars are evenly distributed azimuthally
throughout the Galactic disk peaking radially at 4.5 kpc and falling off
exponentially with distance from the center.  Our model does not assume
the spiral structure of the Galaxy where the pulsars are indeed born. 
The location of most of the detected pulsars is clearly correlated
with the spiral arms of the Milky Way.  Perhaps the difficultly of our
model in reproducing the distance and flux distributions is a result of
the lack of including this strong correlation of the location of pulsars
and the spiral arms.

We believe that the fact that we cannot reproduce the decrease in the
ratio of radio-loud to radio-quiet $\gamma$-ray pulsars for EGRET points
to a significant limitation of our overall model.  The fact that we are
not able to simultaneously account for the detected radio flux and
inferred distance distributions and that the radio pulse profiles do not
correlate with the $\gamma$-ray profiles as those detected by the EGRET,
also suggest that we are not adequately accounting for everything.  The
problem with the pulse profiles perhaps implies that the
parameterization of the ratio of radio core to cone with period in the
ACC model does not apply for $\gamma$-ray pulsars.  Recently Crawford,
Manchester \& Kaspi (2001) reported that six young pulsars have large
linearly polarized profiles characteristic of conal emission.  Yet in
the ACC model, young pulsars with short periods would be characterized by significant core
emission.  In order to avoid the core component, the impact angle must
be fairly large to observe a single peak from the edge of a conal beam.
The pulsar J1105-6107 has a period of 0.063 s, and the ACC model
predicts an integrated flux core to cone ratio of about 60 making it
difficult to avoid seeing the core beam when a significant mean conal
flux of 1.0 mJy is detected.  Perhaps partial conal emission becomes
dominant for short period pulsars as suggested by Manchester (1996).

The correlation between the radio and $\gamma$-ray beam profiles and the
shapes of the profiles are sensitive to the geometry of the beams as
well as the viewing geometry.  In the ACC model that we have used in our
simulations, short period pulsars have fluxes that are core
dominated.  All two-peaked $\gamma$-ray pulse profiles display a single
radio core peak in the radio pulse profile.  In order to see a cone
dominated radio profile, the impact parameter must be so large that
single-peaked $\gamma$-ray profiles are seen.  This
appears contrary to the features of the most of the $\gamma$-ray
profiles detected EGRET and again raises the issue of whether
short period pulsars are core dominated and perhaps are really partial
cone dominated as suggested by Manchester (1996) and, more recently, by
Crawford, Manchester \& Kaspi (2001).

We conclude that a better radio beam model is required to account for
the observed characteristics of radio pulsars.  In addition, recent
theoretical work by Dyks \& Rudak (2003) and Muslimov \& Harding (2004)
indicates the importance of including the caustic component of
$\gamma$-rays in the emission geometry.  We hope that in the near
future, we will be able to include more realistic emission geometries of
the radio and $\gamma$-ray beams.

\section{Acknowledgements} We would like to acknowledge the many
conversations with Zaven Arzoumanian and the insight that he has given
to us. We also thank an anonymous referee for suggesting many 
improvements to the manuscript.  
We appreciate the $S_{min}$ code for the Parkes multi-beam
pulsar survey from Froney Crawford.  We are grateful to Alberto
Pellizzoni in providing us with the all-sky sensitivity maps for AGILE,
as well as Isabelle Grenier in allowing us to use their EGRET all-sky
sensitivity map. We express our gratitude for the generous support of
Research Corporation (CC5813), of the National Science Foundation (REU
and AST-0307365), and NASA Astrophysics Theory Program.


\newpage 

\begin{figure} 
\epsscale{0.75} 
\plotone{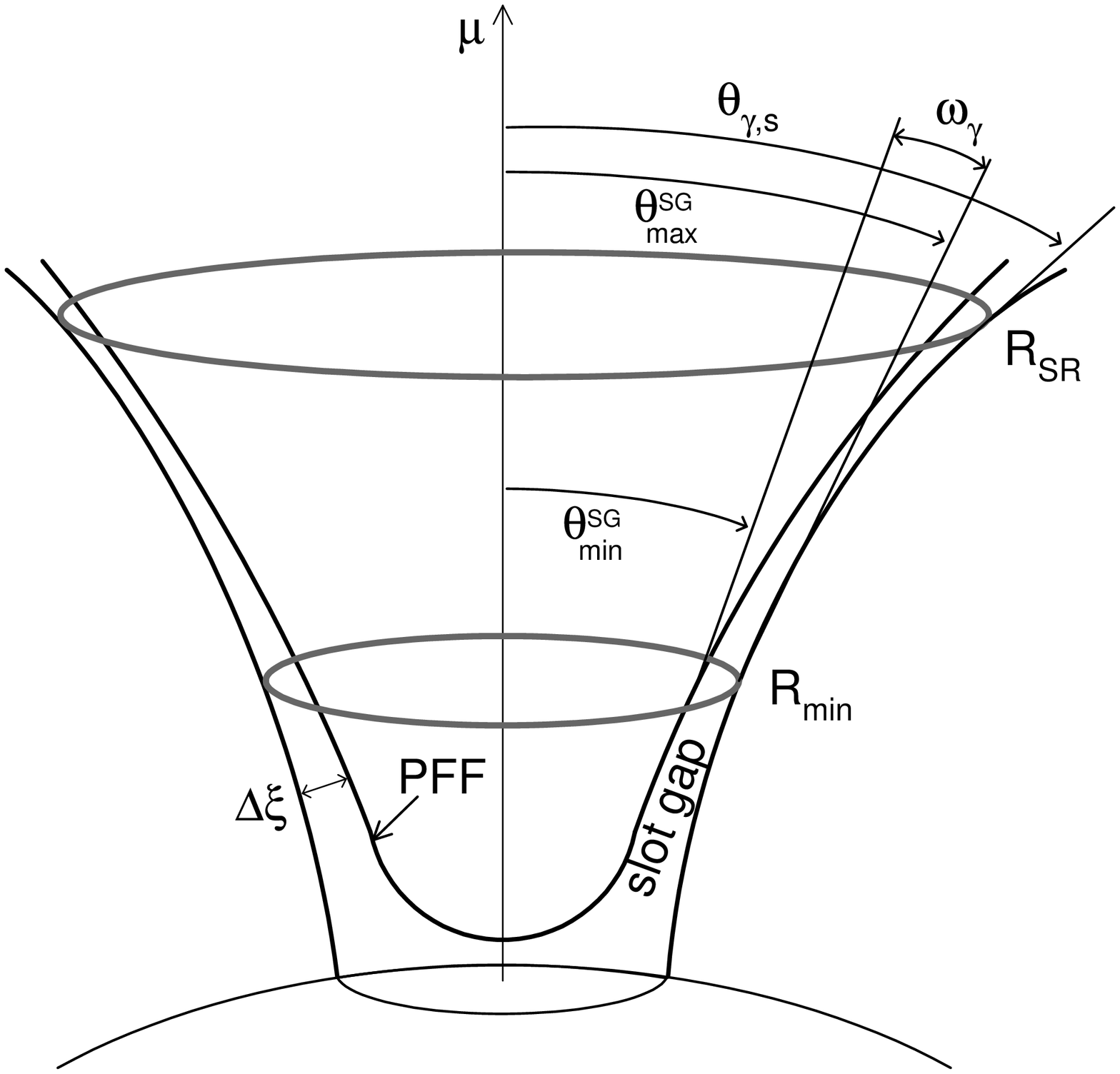} 
\caption{Geometry of the slot gap emission relative to the magnetic
moment vector, $\mu$. The pair formation front (PFF) curves upward near
the last open field line and forms the inner boundary of the slot gap
with width $\Delta\xi$ (in units of polar cap half-angle). $R_{\min}$
and $R_{SR}$ are the minimum and maximum radii of pair synchrotron
radiation. $\theta^{SG}_{\min}$ and $\theta^{SG}_{\max}$ are the tangent
angles to field lines at the inner and outer edge of the slot gap at
$R_{\min}$.}
\end{figure}

\begin{figure} 
\epsscale{0.75} 
\plotone{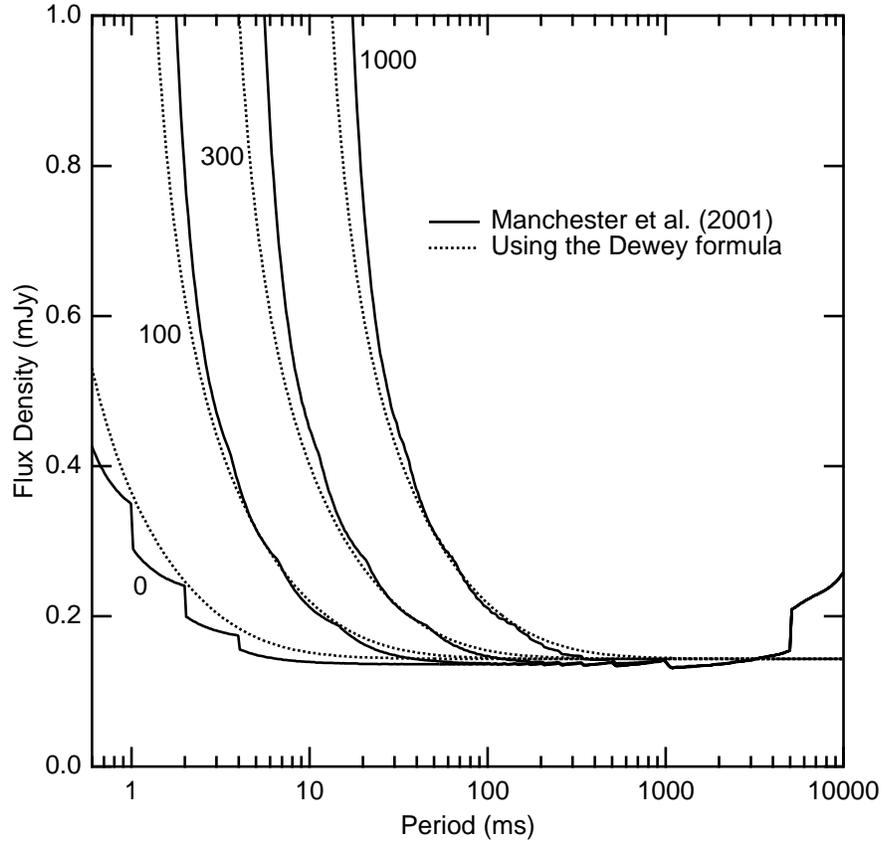} 
\caption{The minimum radio flux, $S_{min}$, detectable by the Parkes
multi-beam pulsar survey for the indicated DM (in $\rm pc\, cm^{-3}$) as
a function of period, calculated as in Manchester et al. (2001) (solid
curves) and calculated using the formula in Dewey et al. (1985) (dotted
curves) with an additional factor of 2 (see text).}
\end{figure}

\begin{figure} 
\epsscale{0.9} 
\plotone{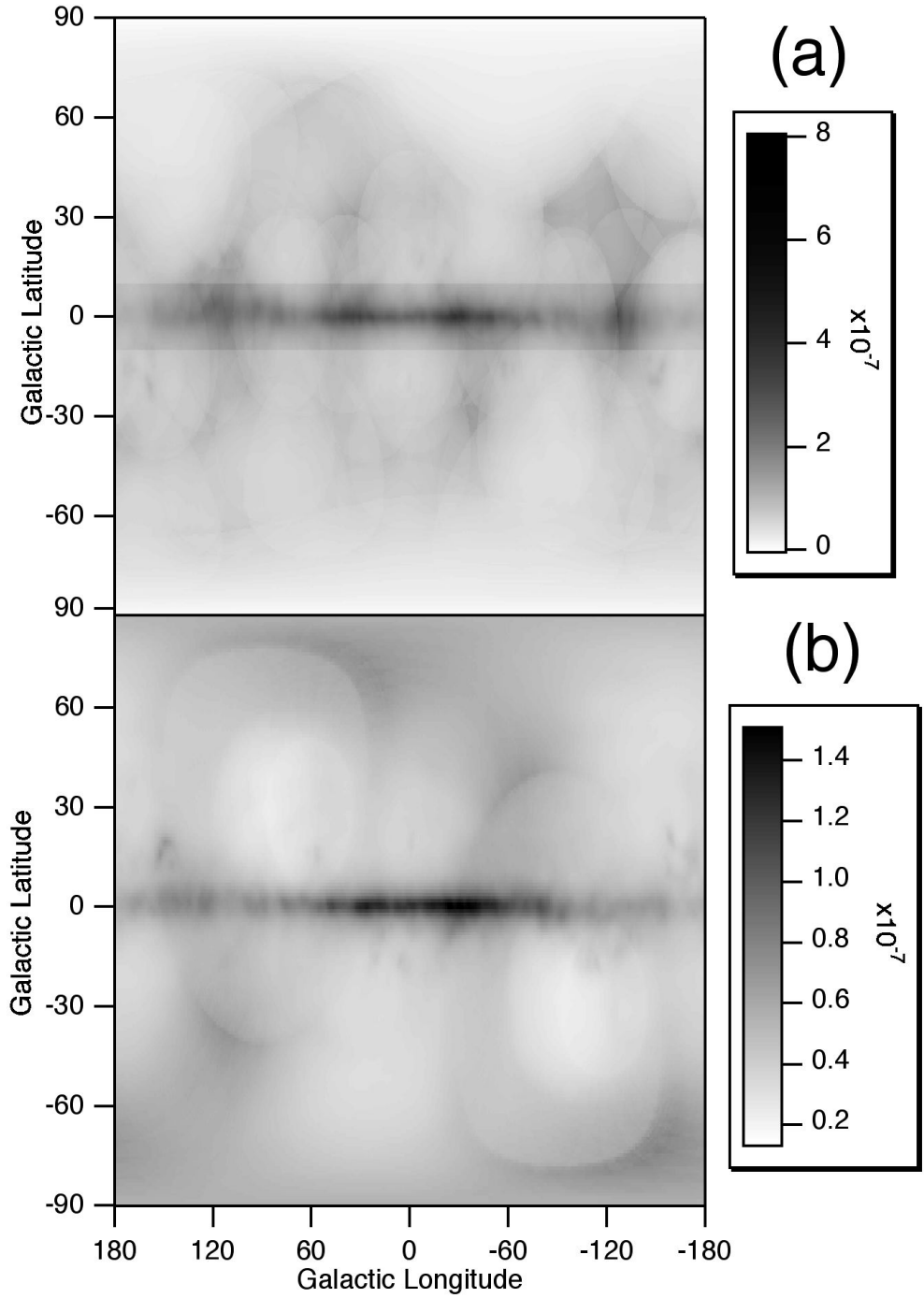} 
\caption{Gamma-ray flux threshold maps in Galactic latitude versus
longitude for EGRET (I. Grenier, private communication) in (a) and for
AGILE (A. Pellizzoni, private communication) in (b) for the nominal case. Gray scale is in
units of photons $\rm s^{-1} cm^{-2}$.}
\end{figure}

\begin{figure} 
\epsscale{0.75} 
\plotone{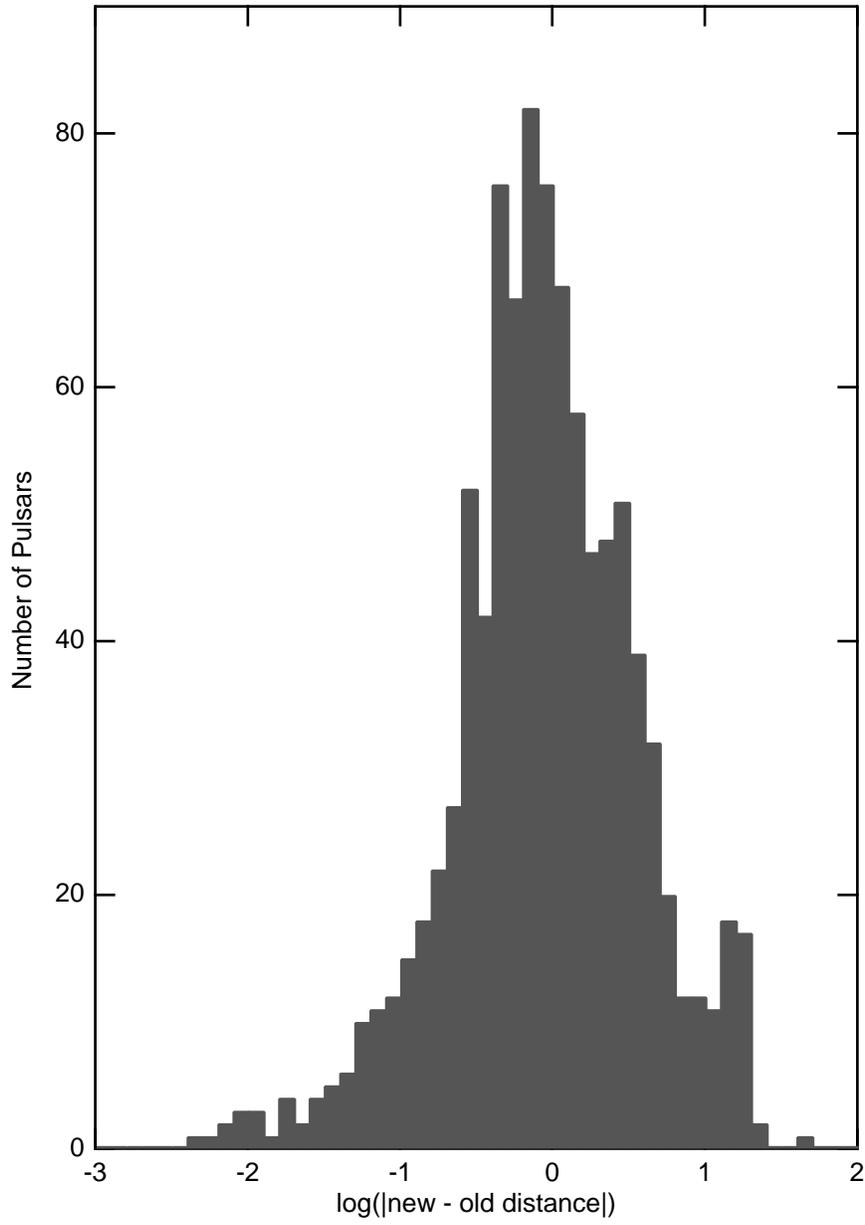} 
\caption{The distribution of the logarithm of absolute value of the
difference between the distance of pulsars obtained from the new (Cordes
\& Lazio 2002) and old (Taylor \& Cordes 1993) electron density models.}
\end{figure}

\begin{figure} 
\epsscale{1.0} 
\plotone{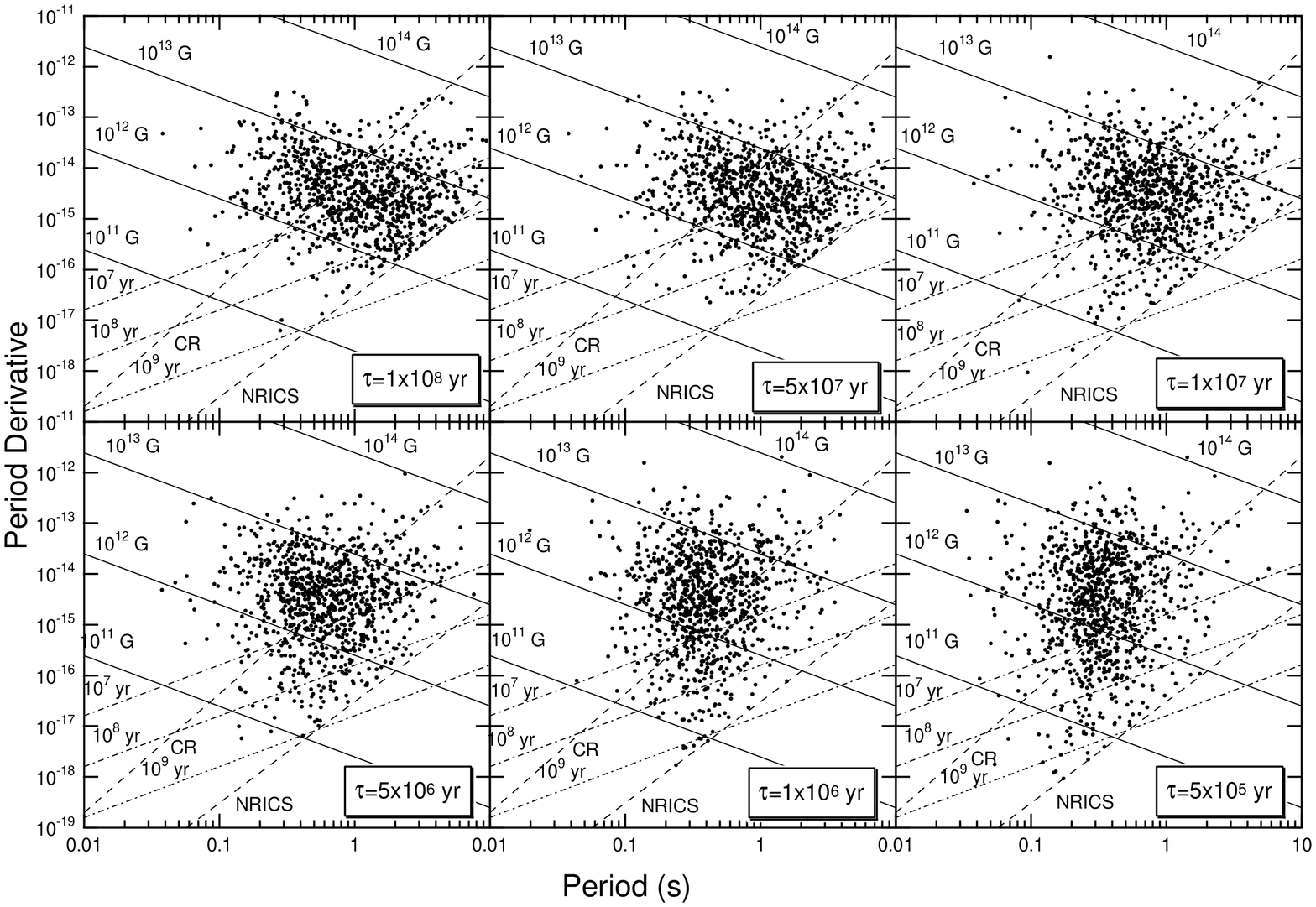} 
\caption{Distributions of simulated radio pulsars as a function of the
period derivative and period for the indicated decay constants of the
magnetic field.  Dashed lines represent the death lines for curvature
radiation (upper) and for nonresonant inverse Compton scattering
(lower). Dot-dashed lines represent the indicated pulsar age (in yr) and
solid lines represent the indicated magnetic surface field strength (in
units of G) assuming a constant dipole spin-down field.}
\end{figure}

\begin{figure} 
\epsscale{1.0} 
\plotone{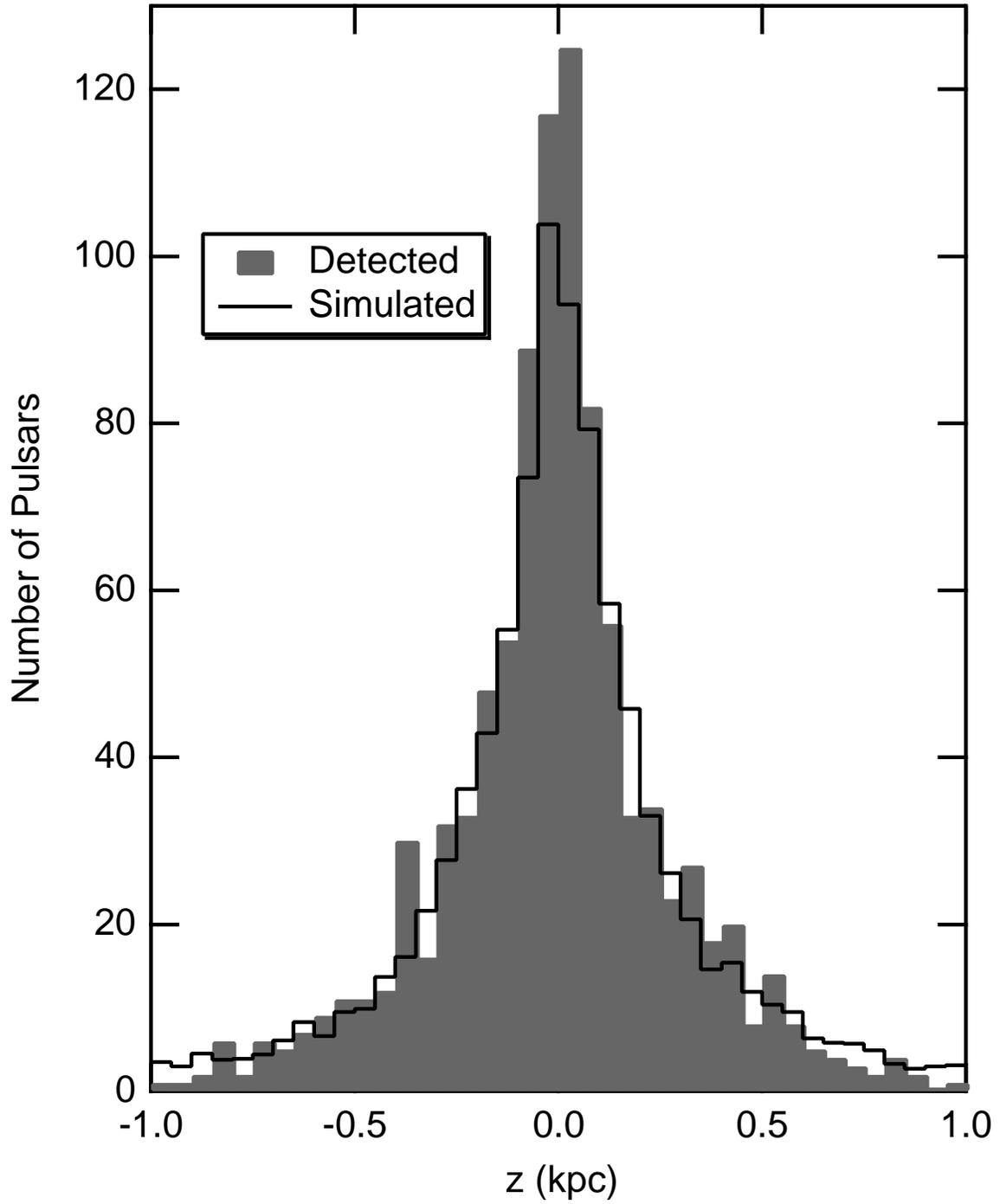} 
\caption{Distribution of the distance z from the Galactic disk for
detected pulsars (shaded histogram) and simulated pulsars (regular
histogram).}
\end{figure}

\begin{figure} 
\epsscale{1.0} 
\plotone{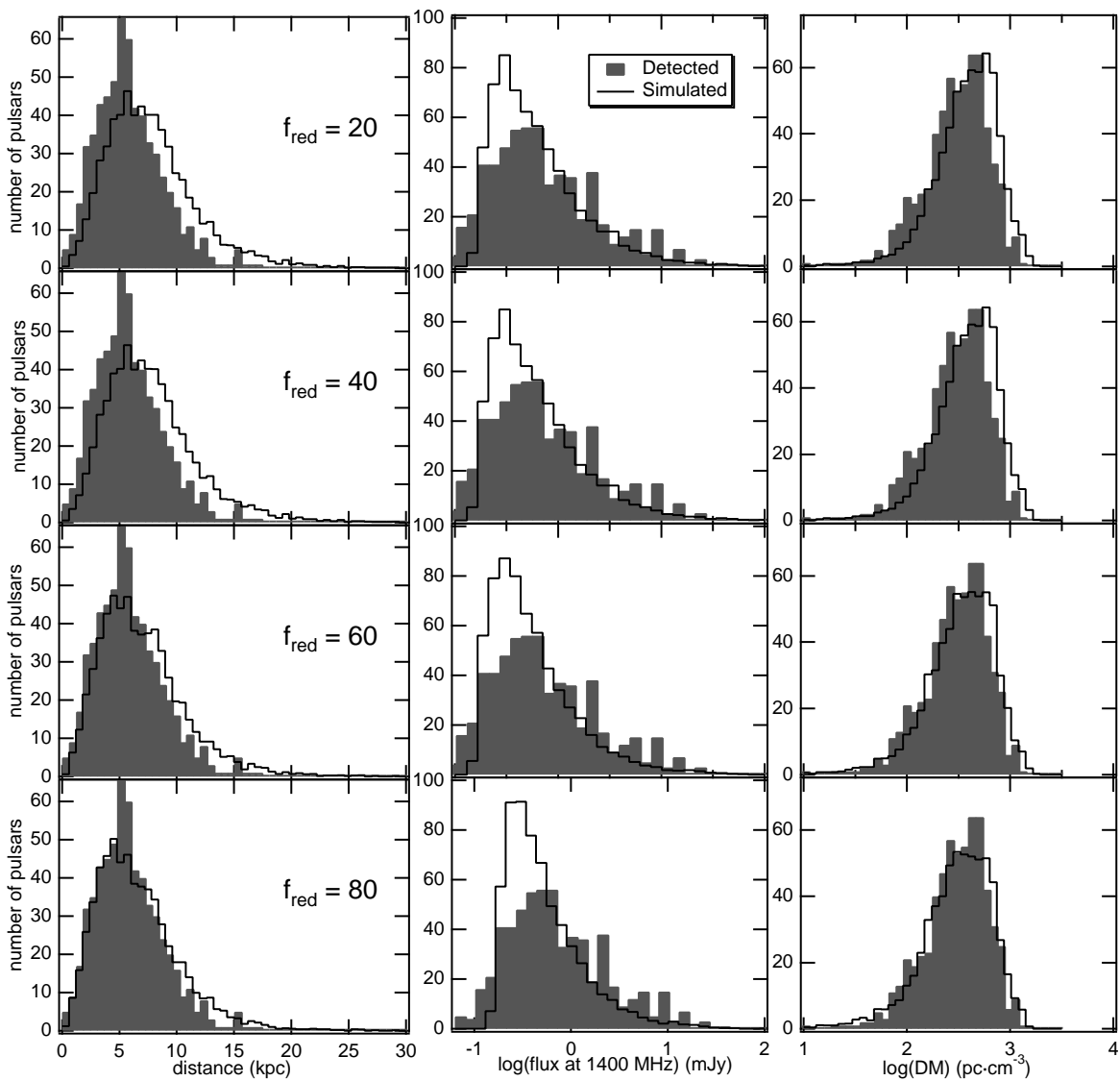} 
\caption{Distributions of the distance, flux at 1400 MHz and dispersion
measure for detected pulsars in the PMBPS (solid histograms) and
simulated pulsars (regular histograms) for the indicated factor,
$f_{\rm red}$, reducing the radio luminosity of the ACC model.}
\end{figure}

\begin{figure} 
\epsscale{1.0} 
\plotone{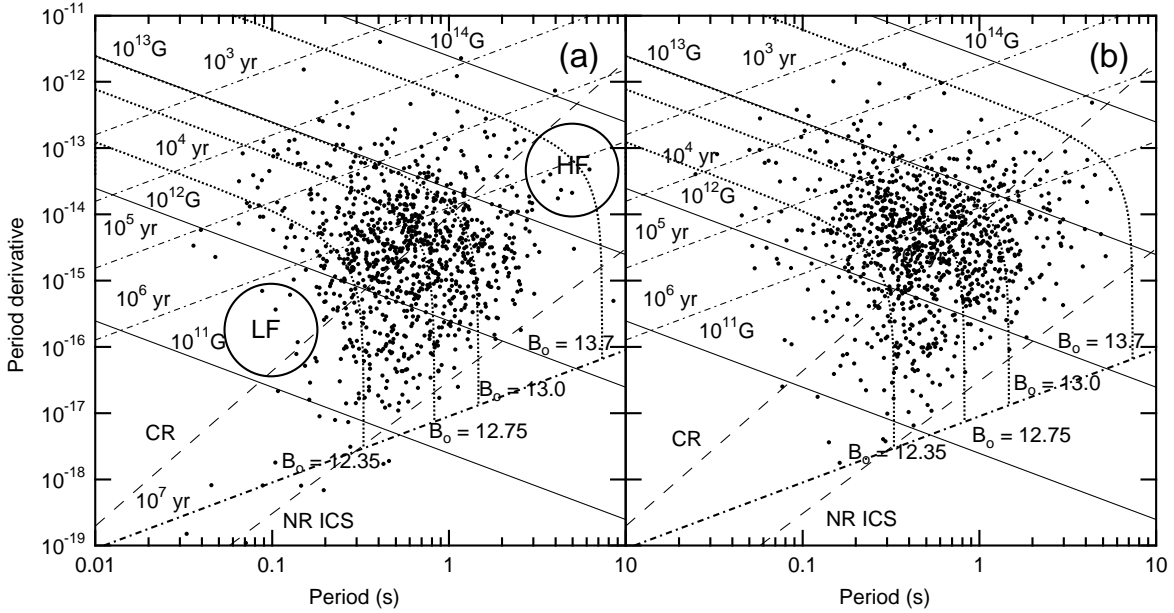} 
\caption{Distributions of (a) detected and (b) simulated pulsars with a
field-decay constant of 2.8 Myr.  Dashed lines represent the death lines
for curvature radiation (upper) and for nonresonant inverse Compton
scattering (lower). Dot-dashed lines represent the indicated pulsar age
assuming a field decay of 2.8 Myr and solid lines represent the
indicated magnetic surface field strength assuming a constant dipole
spin-down field. Dotted curves for the indicated field strengths assume
a decay constant of 2.8 Myr of the magnetic field.  Two regions of
pulsar paucity are indicated by circles - one at low field (LF) and
another at high field (HF).}
\end{figure}

\begin{figure} 
\epsscale{1.} 
\plotone{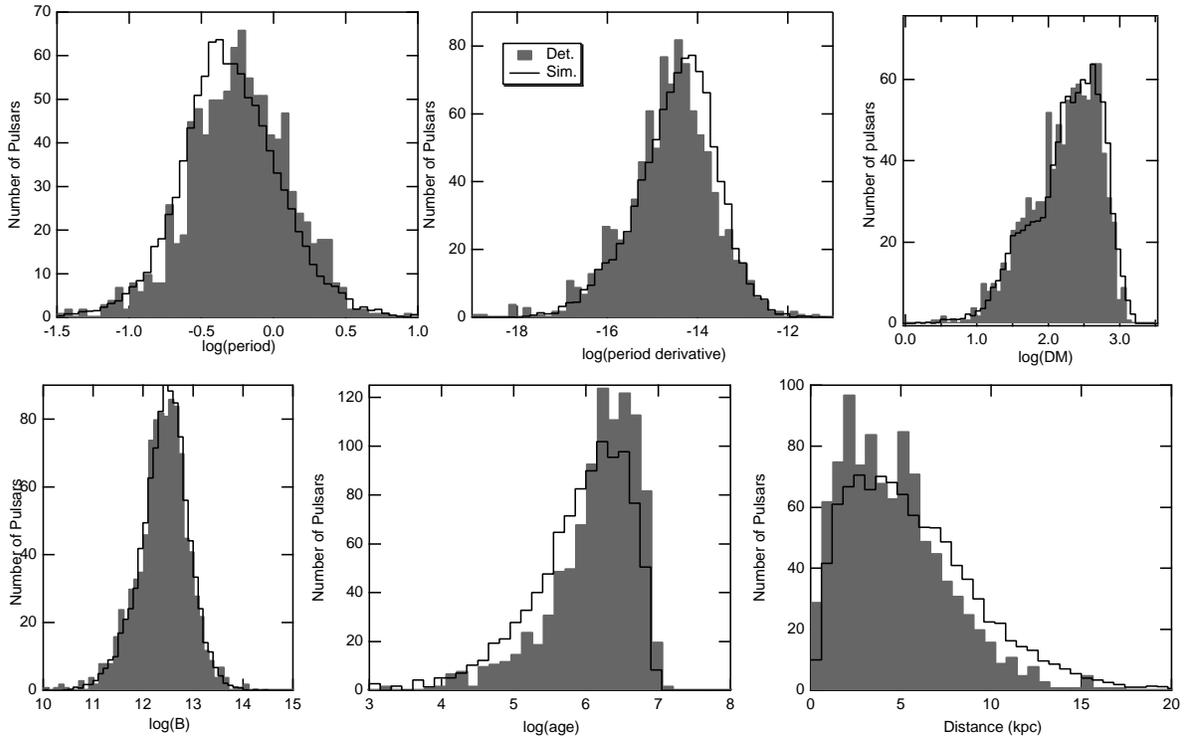} 
\caption{Distributions of various pulsar characteristics indicated as
shaded histograms (detected pulsars) and plain histograms (simulated
pulsars). }
\end{figure}

\begin{figure} 
\epsscale{1.} 
\plotone{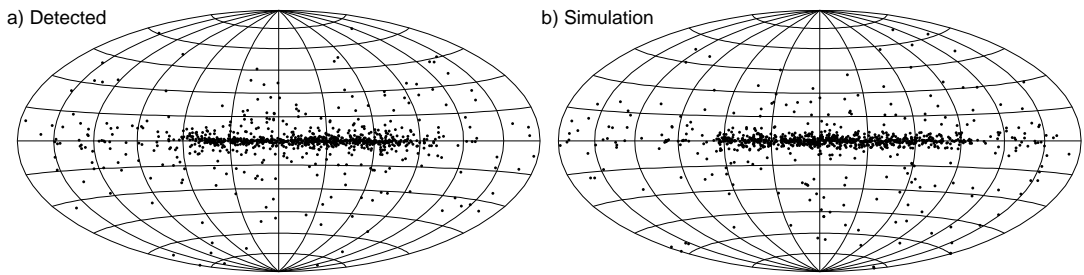} 
\caption{Aitoff plots of the (a) detected and (b) simulated radio
pulsars.}
\end{figure}

\begin{figure} 
\epsscale{1.0} 
\plotone{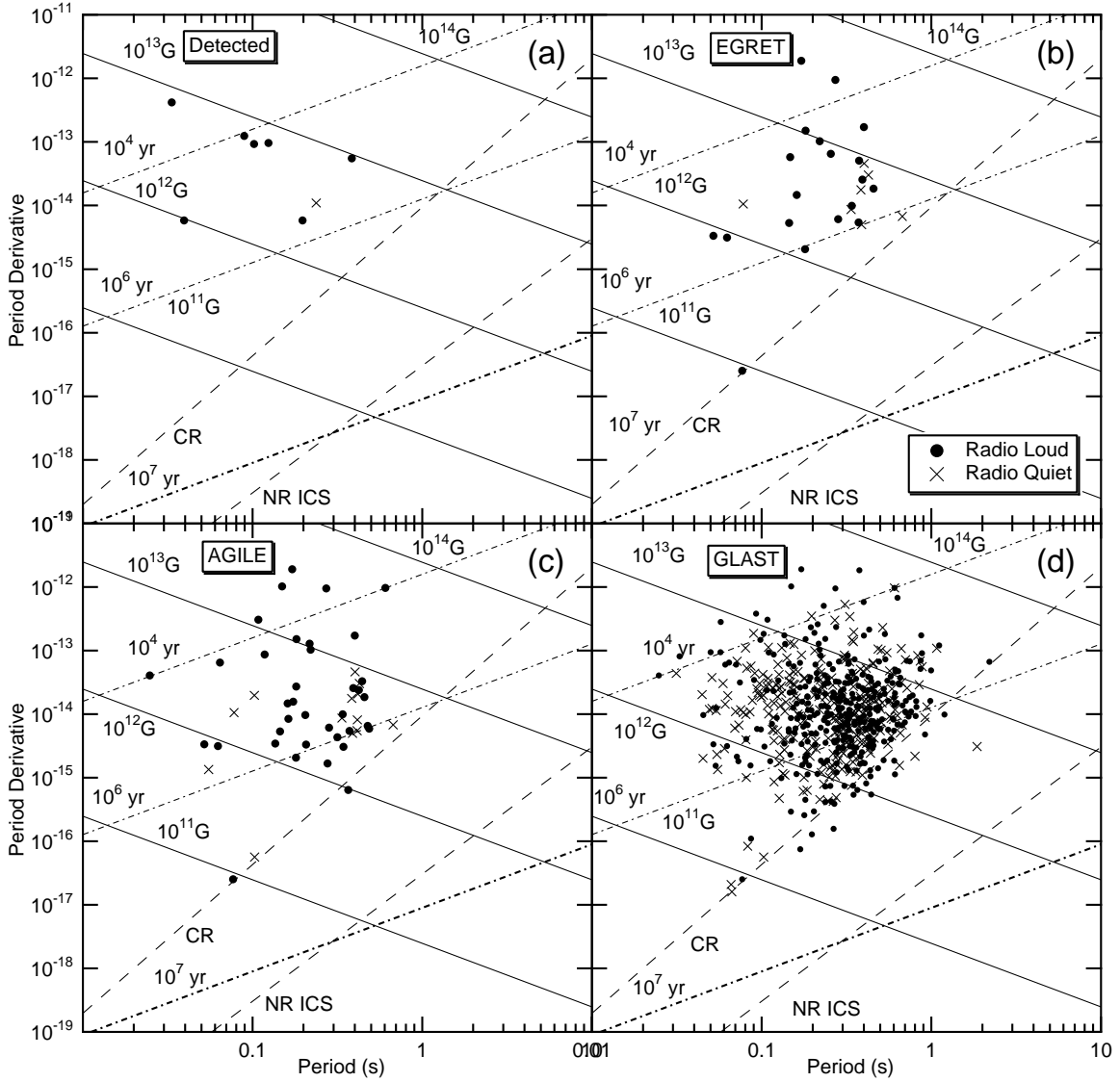} 
\caption{Distributions of radio-quiet (crosses) and radio-loud (solid
dots) $\gamma$-ray pulsars (a) detected by EGRET and simulated for (b)
EGRET (c) AGILE and (d) GLAST, assuming a field-decay constant of 2.8
Myr.  Dashed lines represent the death lines for curvature radiation
(CR) and for nonresonant inverse Compton scattering (NRICS). Dot-dashed
lines represent the indicated pulsar age assuming a field decay of
2.8 Myr and solid lines represent the indicated magnetic surface field
strength assuming a constant dipole spin-down field. }
\end{figure}

\begin{figure} 
\epsscale{1.} 
\plotone{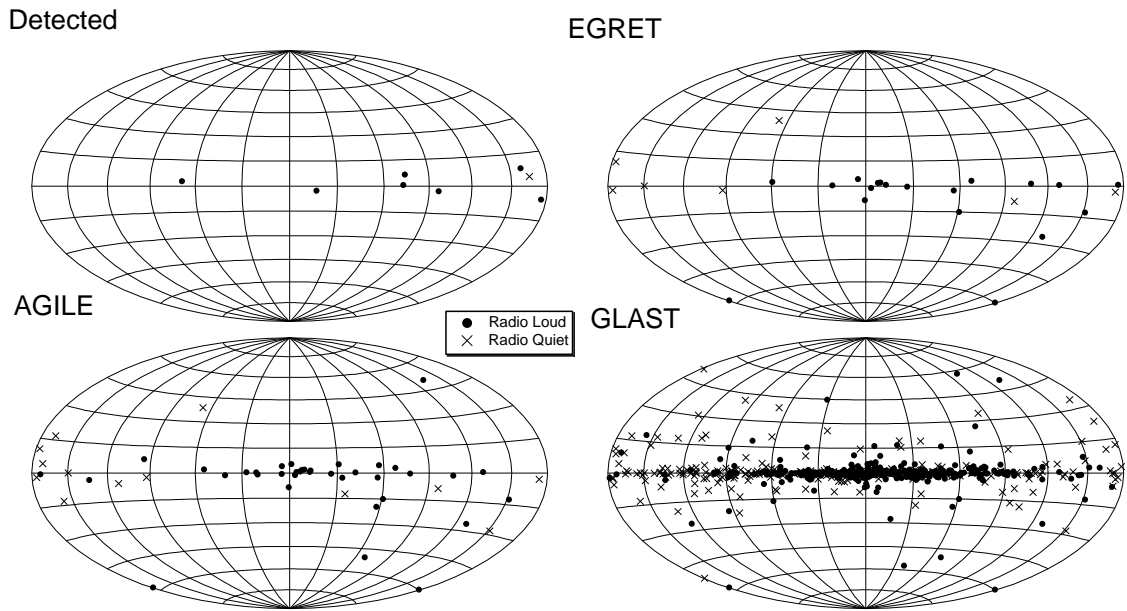} 
\caption{Aitoff plots of radio-quiet (crosses) and radio-loud (solid
dots) $\gamma$-ray pulsars (a) detected by EGRET and simulated for (b)
EGRET (c) AGILE and (d) GLAST, assuming a field-decay constant of 2.8
Myr.}
\end{figure}

\begin{figure} 
\epsscale{0.7} 
\plotone{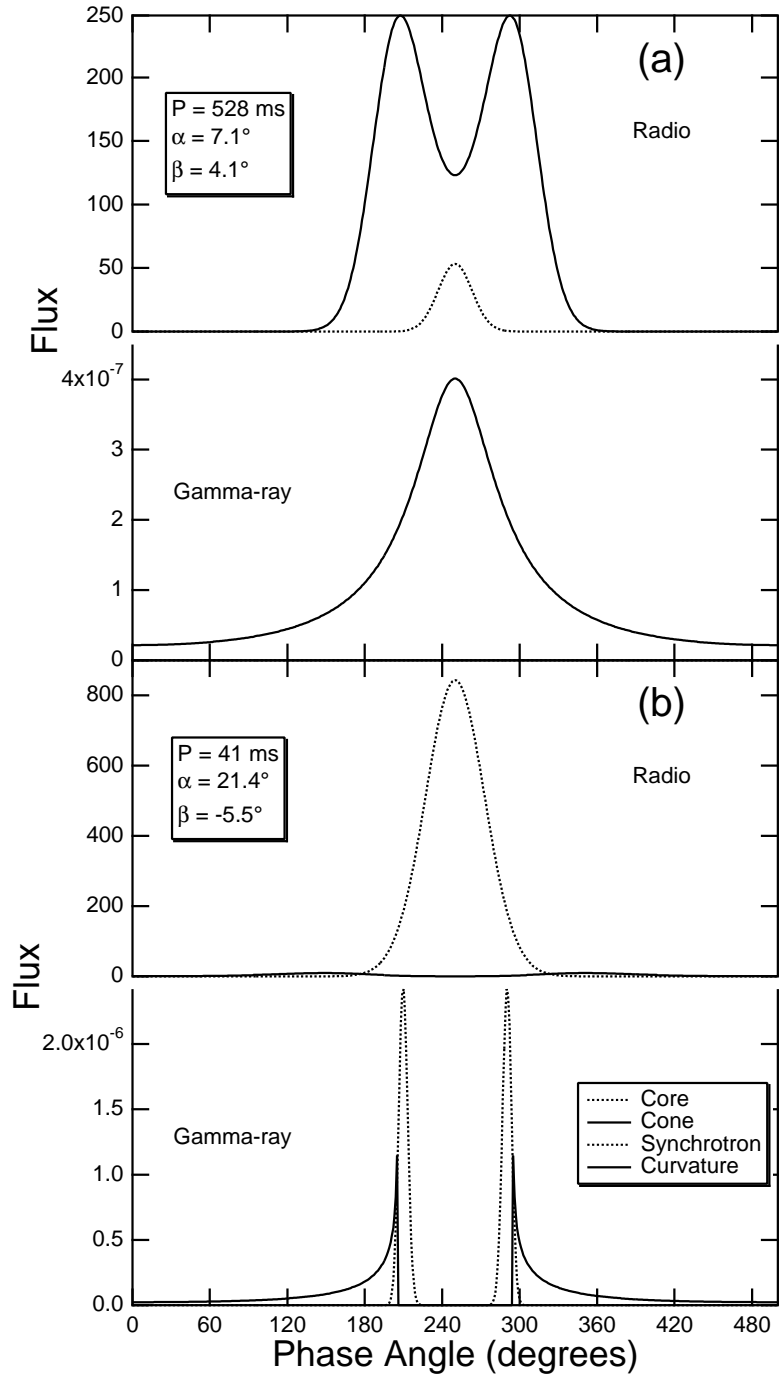} 
\caption{Examples of radio and $\gamma$-ray pulse profiles for two
radio-loud $\gamma$-ray pulsars simulated for EGRET having periods
of 528 ms (a) and 41 ms (b)}.
\end{figure} 
\end{document}